\documentclass[prl,amsmath,amssymb,twocolumn, showpacs, superscriptaddress,10pt]{revtex4-1}

\usepackage{amsmath}
\usepackage{hyperref}
\usepackage{graphicx}
\usepackage{amsfonts}
\usepackage{amsthm}
\usepackage{cases}
\usepackage{bm}
\usepackage{todo}

\usepackage{color}
\definecolor{Blue}{rgb}{0.00, 0.00, 1.00}
\definecolor{Red}{rgb}{1.00, 0.00, 0.00}

\newcommand{\E}{\mathbb E}

\hypersetup{
	colorlinks=true,       
	linkcolor=red,          
	citecolor=blue,        
	filecolor=magenta,      
	urlcolor=cyan           
}

\newcommand{\be}{\begin{equation}}
\newcommand{\ee}{\end{equation}}
\newcommand{\bea}{\begin{eqnarray}}
\newcommand{\eea}{\end{eqnarray}}

\newcommand{\beq}{\begin{equation}}
\newcommand{\eeq}{\end{equation}}
\newcommand{\beqn}{\begin{eqnarray}}
\newcommand{\eeqn}{\end{eqnarray}}

\DeclareMathOperator{\sech}{sech}

\DeclareMathOperator{\sgn}{sgn}

\DeclareMathOperator{\Pf}{Pf}
\DeclareMathOperator{\Tr}{Tr}

\newcommand{\R}{\mathbb{R}}
\newcommand{\C}{\mathbb{C}}
\newcommand{\N}{\mathbb{N}}
\newcommand{\D}{\mathbb{D}}
\newcommand{\Z}{\mathbb{Z}}

\newcommand{\dx}[1]{\ensuremath{\mathrm d} #1}

\newcommand{\e}{{\rm e}}

\begin{document}
	
	\title{Exact persistence exponent for the $2d$-diffusion equation and related Kac polynomials}
	
	\author{Mihail Poplavskyi}
	\affiliation{King's College London, Department of Mathematics, London WC2R 2LS, United Kingdom}
	\author{Gr\'egory \surname{Schehr}}
	\affiliation{LPTMS, CNRS, Univ. Paris-Sud, Universit\'e Paris-Saclay, 91405 Orsay, France}

	\date{\today}
	
	\begin{abstract}
We compute the persistence for the $2d$-diffusion equation with random initial condition, i.e., the probability $p_0(t)$ that the diffusion field, at a given point ${\bf x}$ in the plane, has not changed sign up to time~$t$. For large $t$, we show that $p_0(t) \sim t^{-\theta(2)}$ with $\theta(2) = 3/16$. Using the connection between the $2d$-diffusion equation and Kac random polynomials, we show that the probability $q_0(n)$ that Kac polynomials, of (even) degree $n$, have no real root decays, for large $n$, as $q_0(n) \sim n^{-3/4}$. We obtain this result by using yet another connection with the truncated orthogonal ensemble of random matrices. This allows us to compute various properties of the zero-crossings of the diffusing field, equivalently of the real roots of Kac polynomials. Finally,  we unveil a precise connection with a fourth model: the semi-infinite Ising spin chain with Glauber dynamics at zero temperature.  
\end{abstract}
	

	
	\maketitle
	
Persistence and first-passage properties have attracted a lot of interest during the last decades in physics, both theoretically \cite{Red01,Maj99,BMS13} and experimentally~\cite{TZSS97,WMWC01,DLWCDS02,TS12}, as well as in mathematics~\cite{AS15}. For a stochastic process $X(t)$, the 
persistence $P_0(t)$ is the probability that it has not changed sign  up to time $t$.
In non-equilibrium statistical physics, this is an interesting observable which is non-local in time and  
thus carries useful information on the full history of the system on a given time interval \cite{Wat96}. 

In many physically relevant situations, $P_0(t)$ decays algebraically at late time $t\gg 1$, $P_0(t) \sim t^{-\theta}$, where $\theta$ is called the {\it persistence exponent}~\cite{Red01,Maj99,BMS13}. For instance, for Brownian motion, which is a Markov process, $\theta = 1/2$.
But in many cases, in particular for
coarsening dynamics \cite{DBG94}, and more generally for non-Markov processes, the exponent $\theta$ is non-trivial 
and extremely hard to compute~\cite{Red01,Maj99,BMS13}. Consequently, there are very few non-Markov processes, for which $\theta$ is known exactly. 
One notable example is the $1d$ Ising chain with Glauber dynamics. In this case, at temperature $T=0$, the persistence exponent for the local magnetization can be computed exactly, yielding $\theta_{\rm Ising} = 3/8$~\cite{DHP95,DHP96}. 
   
\begin{figure}[bb]
\includegraphics[width = \linewidth]{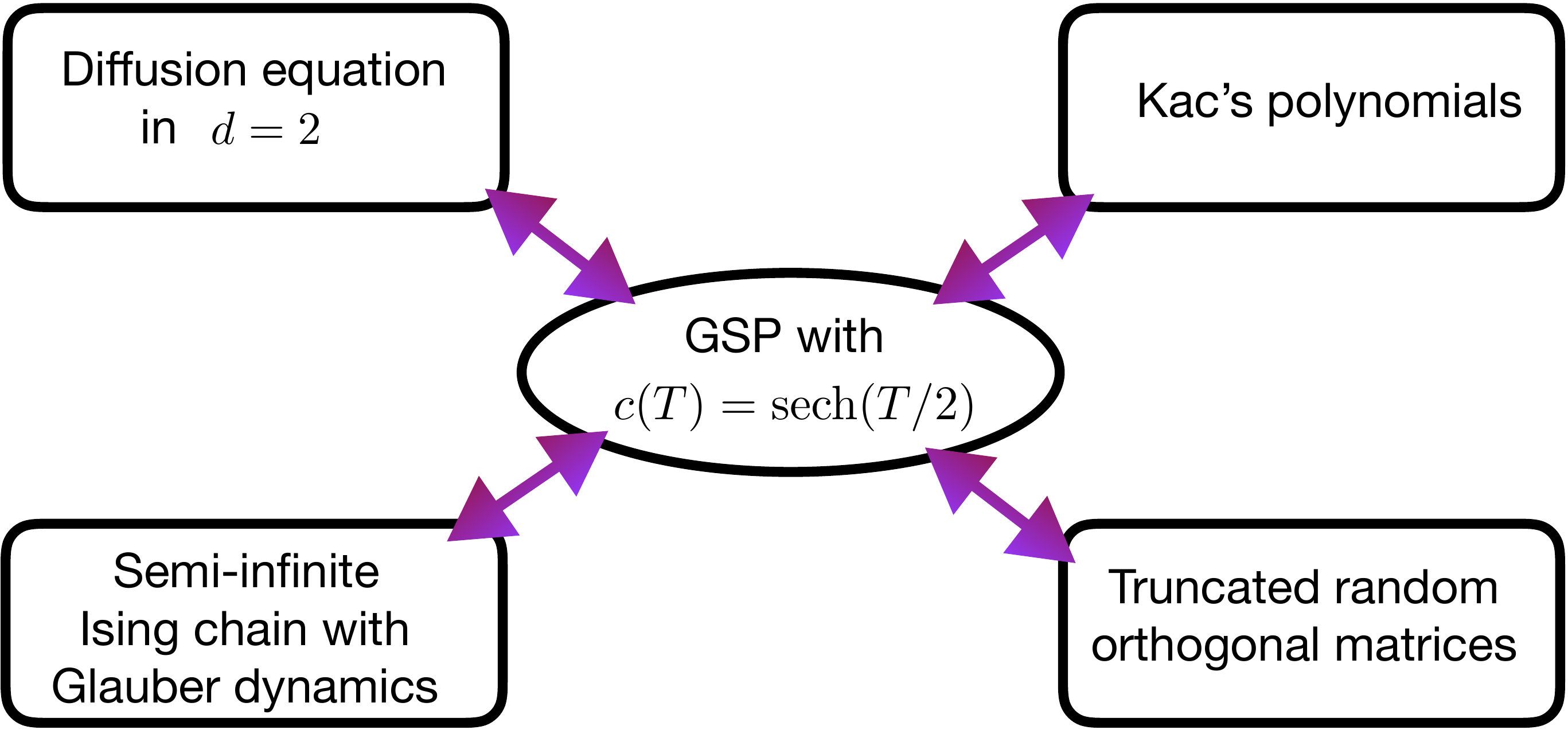}
\caption{Connections between the four models studied here: they are related to 
the same Gaussian stationary process (GSP) with correlator $c(T) = {\rm sech}(T/2)$. Our main results is the exact value of the persistence exponent for this GSP, $\theta(2) = b = 3/16$, together with the full statistics of its zero-crossings (\ref{large_dev})-(\ref{cumulants}).} \label{Fig:recap}
\end{figure}

Another example which has attracted a lot of attention \cite{MSBC96, DHZ96, Hil00,NL01,SM07,SM08,DM15} is the $d$-dimensional diffusion equation 
where the scalar field $\phi({\bf x},t)$ at point ${\bf x} \in {\mathbb R}^d$ and time $t$ evolves as $\partial_t \phi({\bf x},t) = \Delta \phi({\bf x},t)$
where initially $\phi({\bf x},t=0)$ is a Gaussian random field, with zero mean and short range correlations $\langle \phi({\bf x},0) \phi({\bf x'},0)\rangle = \delta^{d}({\bf x} - {\bf x'})$. 
For a system of linear size $L$, the persistence $p_0(t,L)$ is the probability that $\phi({\bf x},t)$, at some fixed point {\bf x} in space, does not change sign up to time $t$ \cite{MSBC96,DHZ96}. We assume that ${\bf x}$ is far enough from the boundary, where  
the system is invariant under translations, and $p_0(t,L)$ is thus independent of ${\bf x}$. It was shown \cite{MSBC96,DHZ96} that $p_0(t,L)$ takes the scaling form, for large $t$ and large $L$, with $t/L^2$ fixed
\begin{eqnarray}\label{scaling}
p_0(t,L) \sim L^{-2 \theta(d)} h(L^2/t) \;,
\end{eqnarray}
with $h(u) \to c_1$, a constant, when $u \to 0$ and $h(u) \propto u^{\theta(d)}$ when $u \to \infty$ where $\theta(d)$ was found, numerically, to be non-trivial, e.g. $\theta(1) = 0.1207\ldots$, $\theta(2) = 0.1875\ldots$ \cite{MSBC96,DHZ96,NL01}. This scaling form (\ref{scaling}) shows that $p_0(t,L) \sim t^{-\theta(d)}$ for an infinite system. Alternatively, $\theta(d)$ can also be obtained, in a finite system of size $L$, from $p_0(t,L) \sim L^{-2 \theta(d)}$ for $t \gg L^2$. To study $p_0(t,L)$ it is useful to introduce the normalised process $X(t) = {\phi({\bf 0},t)}/{\langle \phi({\bf 0},t)^2\rangle}$~\cite{foot_otherpoint}. Being Gaussian, $X(t)$ is completely characterised by its autocorrelation function which, for an infinite system $L \to \infty$, behaves like $C(t,t') = \langle X(t) X(t')\rangle= (2\sqrt{t\,t'}/(t+t'))^{d/2}$. In terms of logarithmic time $T= \ln t$, $Y(T) = X(\e^T)$ is a Gaussian stationary process with covariance $c(T) = [{\rm sech}(T/2)]^{d/2}$ (see Fig. \ref{Fig:recap}). In particular, $c(T) \approx 1 - d\,T^2/16$ for $T \to 0$, indicating a smooth process with a finite density of zero-crossings $\rho_0= (2\pi)^{-1}\sqrt{d/2}$ \cite{Ric44}. Although several very accurate approximation schemes exist to compute $\theta(d)$~\cite{MSBC96,DHZ96,MB98,EMB04,Hil00}, there is not a single value of $d$ for which this persistence exponent could be computed exactly. 
In this Letter, we focus on the case $d=2$. As we show below, this case 
is particularly interesting because it is related to a variety of other interesting models (see Fig. \ref{Fig:recap}), in particular to the celebrated Kac's random polynomials \cite{DPSZ02,SM07,SM08,DM15}. These are polynomials of degree~$n$
\begin{equation}
K_n(x) = \sum_{i=0}^n a_i \, x^i \;, \label{def_Kac}
\end{equation}   
where the coefficients $a_i$'s are independent and identically distributed (i.i.d.) real Gaussian random variables of zero mean and unit variance.  
Of course, $K_n(x)$ has $n$ roots in the complex plane. These roots tend to cluster, when $n \to \infty$, close to the unit circle centered at 0. But because the coefficients $a_i$'s  are real, the statistics of the number of real roots is singular, and it has thus generated a lot of interest~\cite{EK95,BD97}. In particular, the average number of real roots grows, for $n \gg 1$, like $\sim (2/\pi) \ln n$, hence much smaller than $n$.  
It is thus natural to ask: what is the probability $q_0(n)$ that $K_n(x)$ has no real roots
for an even $n$? It was shown in Ref. \cite{DPSZ02} that $q_0(n)$ decays to zero as 
$q_0(n) \sim n^{- 4 b}$
where, remarkably, $b$ turns out to be the persistence exponent for the diffusion equation in $d=2$ \cite{SM07,SM08}, i.e. $b = \theta(2)$. 

To establish the connection between these two problems, one first notices that almost all the real roots of $K_n(x)$ lie very close to $\pm 1$, in a window of size ${\cal O}(1/n)$~\cite{AF04}. In addition, one can show that the real roots of $K_n(x)$ behave independently and identically within each of the 
four sub-intervals $(-\infty,-1]$, $[-1,0]$, $[0,1]$ and $[1, + \infty)$. One can thus focus on one of these intervals, say $[0,1]$, and consider $\tilde q_0(x,n)$, which is the probability that $K_n$ has no real root in $[0,x]$, with $0<x\leq1$. Clearly, $q_0(n) \sim [\tilde q_0(1,n)]^4$ for large $n$. For $x \to 1^-$, it was shown in \cite{SM07,SM08,DM15,DPSZ02} that the behavior of $\tilde q_0(x,n)$ is governed by 
the zero-crossings properties of the GSP $Y(T)$ with covariance $c(T) =  {\rm sech}\,{(T/2)}$, i.e., the same GSP that 
governs the zero-crossings of the $2d$-diffusion equation (see Fig. \ref{Fig:recap}). In particular, in the scaling limit $n \to \infty$, $x \to 1^-$ with $n\,(1-x)$ fixed (recall that the scaling region around $\pm 1$ is of order ${\cal O}(1/n)$), one can show that $\tilde q_0(x,n)$ takes the scaling form \cite{SM07, SM08}, 
\bea\label{fss_poly}
\tilde q_0(x,n) \sim n^{-b} \tilde h\,(n(1-x)) \;,
\eea
with {$\tilde h(u) \to c_2$}, a constant, for $u \to 0$ and $\tilde h(u) \sim u^{b}$ for $u \to \infty$. The large $u$ behaviour follows from the fact that $\tilde q_0(x, n \to \infty)$ is well defined. This form (\ref{fss_poly}) is the
exact analogue of the finite size scaling form in Eq.~(\ref{scaling}), with $n$ playing the role of $L^2$ and $(1-x)$ the role of inverse time $1/t$~\cite{SM07, SM08}. This implies that $b=\theta(2)$ can be extracted either for finite $n$, from  $\tilde q_0(1,n) \sim n^{-b}$, or for $n \to \infty$ (i.e. for the Gaussian power series) from $\tilde q_0(x, n \to \infty) \sim (1-x)^{b}$, as $x \to 1^-$. The study of this exponent $b$ has generated a lot of interest in the maths literature \cite{AS15,DM15,DPSZ02,LS02,LS05,Mol12} and the best existing bounds are $0.144338\ldots = 1/(4 \sqrt{3}) \leq b \leq 1/4=0.25$ \cite{LS02, Mol12}.

{\it Main results.} Here, we exploit a connection between the Kac's polynomials and the so-called truncated real orthogonal ensemble of random matrices~\cite{Kr09,Fo10,KSZ10} (see below) to obtain the exact result
\begin{equation}\label{exact_theta_2}
b=\theta(2) = {3}/{16} = 0.1875 \;,
\end{equation}
which is fully consistent with numerical simulations \cite{MSBC96,DHZ96,NL01} and the above exact bounds \cite{LS02,Mol12} as well as with a recent conjecture in number theory \cite{CH2017}. We also compute the full probability distribution of 
the number of zero-crossings $N_t$ of $\phi({\bf 0},t)$ up to time $t$. Let $p_k(t,L) = {\rm Prob.}(N_t = k)$ and $p_k(t) = p_k(t,L\to \infty)$. We show that, for large $t$ and $k$, with $k/\ln t$ fixed, $p_k(t)$ takes the large deviation form proposed in~\cite{SM07,SM08}
\begin{eqnarray} \label{large_dev}
p_k(t) \sim t^{- \varphi(k/\ln t)} \;,
\end{eqnarray}
where the large deviation function $\varphi(x)$ is computed exactly. Its asymptotic behaviours are given by
\begin{eqnarray}\label{asym_phi}
\varphi(x) \sim
\begin{cases}
&\frac{3}{16} + x\, \ln x  \;, \; x \to 0 \\
& \frac{1}{2 \sigma^2}(x - \frac{1}{2 \pi})^2 \;, \; |x-\frac{1}{2 \pi}| \ll 1 \\
& \frac{\pi^2}{4}\,x^2 - \frac{\ln 2}{2} \, x \;, x \to \infty 
\end{cases}
\end{eqnarray}
with $\sigma^2 = 1/\pi - 2/\pi^2$. Close to the center, for $x = \frac{1}{2\pi}$, the quadratic behaviour in Eq. (\ref{asym_phi}) shows that $p_k(t)$ has a Gaussian peak, of width $\sigma \ln(t)$, close to its maximum $\langle N_t \rangle \approx \ln(t)/(2 \pi)$. However, away from this central Gaussian regime, $p_k(t)$ is flanked, on both sides of $\langle N_t\rangle$, by non trivial tails (\ref{asym_phi}) -- the right one being however still Gaussian (at leading order), though different from the Gaussian central part. Finally, we also obtain
the large $t$ behavior of the cumulants $\langle N_t^p\rangle_c$ of arbitrary order~$p$ 
\begin{equation}\label{cumulants}
\langle N_t^p\rangle_c \sim \kappa_p \ln t \;,\, \kappa_p = \frac{2^{p-2}}{\pi^{2}} \sum_{m=1}^p (-2)^{m-1}\Gamma^2\left(\frac{m}{2}\right)  {\cal S}_p^{(m)} \;,
\end{equation}   
where ${\cal S}_p^{(m)}$ is the Stirling number of the second kind~\cite{Stirling}. In particular one recovers $\kappa_1 = 1/(2\pi)$ and $\kappa_2 = \sigma^2 = 1/\pi - 2/\pi^2$ (see Ref. \cite{MB98}) and obtains for instance $\kappa_3 = 4/\pi - 12/\pi^2$. 
Our main results in Eqs. (\ref{exact_theta_2}), (\ref{asym_phi}) and (\ref{cumulants}) are not only relevant for the $2d$-diffusion equation, but also for the whole class of models discussed in this Letter that can me mapped onto the GSP, $Y(T)$, with correlator $c(T) = {\rm sech}(T/2)$ (see Fig. \ref{Fig:recap}). In particular, the probability ${\cal P}_k(T)$ that it has exactly $k$ zeros up to $T$ is given, for large $T$ and $k = {\cal O}(T)$, by ${\cal P}_k(T) \sim e^{-T \varphi(k/T)}$, with the same function $\varphi(x)$ (\ref{asym_phi}). Similarly, the cumulants of the number of zero crossings are given by (\ref{cumulants}), with the substitution $\ln t \to T$ and the same coefficients $\kappa_p$. We further show that this GSP has a Pfaffian structure: the multi-time correlation functions of $\sgn(Y(T))$ can be written as Pfaffians \cite{SM}. Besides, we demonstrate that the zeros of $Y(T)$ form a Pfaffian point process \cite{SM}. Finally, we establish an exact mapping between the $2d$-diffusing field and the semi-infinite Ising spin chain with Glauber dynamics at zero temperature. As we will see, using the exact result for the persistence exponent of the full chain, $\theta_{\rm Ising} = 3/8$ \cite{DHP95, DHP96}, this connection provides an alternative derivation of the exact result $\theta(2) = b = \theta_{\rm Ising}/2 = 3/16$.

{\it Truncated random orthogonal matrices}. 
We consider the set of real orthogonal matrices, of size $(2n+1)\times(2n+1)$ (with $n$ a positive integer), uniformly distributed,
with the Haar measure, on the orthogonal group $O(2n+1)$. Let $O$ be such 
a real random orthogonal matrix, such that $O\,O^T = {\mathbb{I}}$. We define its truncation
$M_{2n}$ as the $2n\times 2n$ random matrix obtained by removing the last column and row from the matrix $O$
\begin{equation}\label{truncation}
	O = \begin{pmatrix}
		M_{2n} & \mathbf{u} \\
		\mathbf{v}^T & a
	\end{pmatrix},
\end{equation}
where $\mathbf{u},\mathbf{v}$ are column vectors and $a$ is a scalar. Such truncated
matrices, together with their unitary counterpart, were studied in the context of mesoscopic physics~\cite{ZS00,KSZ10} and extreme statistics~\cite{LGMS18}. The orthogonality condition $O\,O^T = {\mathbb{I}}$ implies that
$M_{2n} M_{2n}^T = {\mathbb{I}} - {\bf u}\, {\bf u}^T$ and hence all the eigenvalues of $M_{2n}$ lie in the unit disk (since their
norm is less than unity). They are the roots of the 
characteristic polynomial $g_M(z) = \det(z{\mathbb I}-M_{2n})$, which after some manipulations, can be written as \cite[Lemma 6.7.2]{HKPV09} (see also \cite{SM})
\begin{equation}\label{identity_det}
	g_M(z) = \det O \det(zM_{2n}-{\mathbb I})\, (a+z\mathbf{v}^T({\mathbb I}-zM_{2n})^{-1}\mathbf{u})\;.
\end{equation}
Since the eigenvalues $z_i$'s of $M_{2n}$ 
are such that $0<|z_i| < 1$, one has necessarily that $\det(z_i\,M_{2n}-{\mathbb I}) = z_i^N g_M(1/z_i)\neq 0$ in the 
right hand side of Eq. (\ref{identity_det}). This implies that the $z_i$'s are the zeros of $(a+z\mathbf{v}^T({\mathbb I}-zM_{2n})^{-1}\mathbf{u})$ [see Eq.~(\ref{identity_det})]. Expanding in powers of $z$ shows that the eigenvalues of $M_{2n}$ are the zeroes of the series 
\bea\label{series_f}
F_{2n}(z) = a+\sum_{k=1}^{\infty}z^k \mathbf{v}^TM_{2n}^{k-1}\mathbf{u} \;,
\eea
with $|z|<1$ (note that $\mathbf{v}^TM_{2n}^{k-1}\mathbf{u}$ are real numbers). Quite remarkably, one can show \cite{HKPV09} that the scaled sequence of the real coefficients of the series in Eq.~(\ref{series_f}), i.e.,
$\sqrt{2n}\{a, \mathbf{v}^T\,\mathbf{u}, \mathbf{v}^T\,M_{2n}\,\mathbf{u}, \mathbf{v}^TM_{2n}^{2}\,\mathbf{u}, \cdots\}$, converges, as $n \to \infty$,  to a sequence of i.i.d. Gaussian random variables, with zero mean and unit variance. 
This implies that, for $n \to \infty$, the real eigenvalues of $M_{2n}$ in (\ref{truncation}) and the real zeroes of $K_n(x)$ in (\ref{def_Kac}) in the interval $[-1,1]$ share the same statistics.

But what about the connection between these two models for finite $n$? In fact, it is known that the eigenvalues of $M_{2n}$ accumulate close to $x = {\pm} 1$, also on a window of size ${\cal O}(1/n)$ \cite{KSZ10}, like for the Kac's polynomials \cite{AF04}. Hence, if one considers the probability $\tilde Q_0(x,n)$ that $M_{2n}$ has no real eigenvalue in $[0,x]$, it is natural to expect that, as for Kac's polynomials (\ref{fss_poly}), for large $n$ and $x \to 1$ keeping $n(1-x)$ fixed, $\tilde Q_0(x,n)$ behaves as
\bea\label{fss_matrix}
\tilde Q_0(x,n) \sim n^{-\gamma}  \tilde H(n(1-x))  \;,
\eea
where the exponent $\gamma$ is yet unknown and, a priori, the scaling function $\tilde H(u)$ is different for $\tilde h(u)$ in Eq.~(\ref{fss_poly}).
However, for $n \to \infty$, we have seen that $\tilde q_0(x,n)$ and $\tilde Q_0(x,n)$ do coincide, since they both correspond to the probability that the (infinite) Gaussian power series has no real root in $[0,x]$. This implies that $\tilde Q_0(x,n \to \infty) = \tilde q_0(x,n \to \infty) \sim (1-x)^{b}$, which, together with the scaling form (\ref{fss_matrix}), shows that $\gamma = b$. Finally, since we expect that $\tilde Q_0(1,n)$ exists, one has 
$\tilde H(u) \to c_3$, a constant, when $u \to 0$, and therefore 
$\tilde Q_0(1,n) \sim n^{-b}$ for large $n$. One can also consider the probability $Q_0(x,n)$ that $M_{2n}$ has no real eigenvalue in $[-x,x]$. Using the statistical independence of the positive and negative real eigenvalues for large $n$, one has $Q_0(x,n) \sim [\tilde Q_0(x,n)]^2$, and in particular $Q_0(1,n) \sim n^{-2 b}$ for large $n$. Using similar arguments, one can show that the full statistics of the zero-crossings of the diffusion equation (equivalently of the real roots of $K_n(x)$) can be obtained, at leading order for large $n$, from the statistics of the number of real eigenvalues ${\cal N}_n$ of the random matrix $M_{2n}$, which we now study. Our analysis follows the line developed in \cite{KPTTZ16} where the real eigenvalues of real Ginibre matrices were studied.


We start with the full joint distribution of the eigenvalues of $M_{2n}$~(\ref{truncation}).  
Since $M_{2n}$ is real and of even size $2n$, it has $l$ (with $l$ even) real eigenvalues (and possibly $l=0$), denoted
by $\lambda_1\leq\ldots\leq \lambda_l$, and $m=n-l/2$ 
pairs of complex conjugate eigenvalues $z_1=x_1+iy_1,z_2=\overline{z_1},\ldots,
z_{2m-1}=x_m+iy_m,z_{2m}=\overline{z_{2m-1}}$ with $x_1\leq\ldots\leq x_m$. Then the ordered eigenvalues of $M_{2n}$ conditioned to have
$l$ real eigenvalues have the joint distribution \cite{KSZ10,Fo10}
	\begin{eqnarray}\label{e:jpdf}
		p^{(l,m)}\left(\vec{\lambda},\vec{z}\right) = 
		C
		\left|\Delta\left(\vec{\lambda},\vec{z}\right)\right|
		\prod_{j=1}^l w\left(\lambda_j\right)
		\prod_{j=1}^{2m} w\left(z_j\right),
	\end{eqnarray}
where $C \equiv C_{m,n}$ is a normalization constant, 
\begin{equation}\label{e:weight}
		w^2(z) = ({2\pi|1-z^2|})^{-1} \;,
\end{equation}
and $\Delta$ is a Vandermonde determinant. The generating function (GF) of ${\cal N}_n$ (the number of real roots) reads
\bea\label{gen_func}
\langle e^{s {\cal N}_n} \rangle_{M_{2n}} = \left \langle \prod_{i=1}^{2n}  1 - (1 - e^{s}) \chi_{\mathbb{R}}(\zeta_i)\large \right \rangle_{M_{2n}} \;,
\eea		
for $s<0$, where the product runs over all the eigenvalues $\zeta_i$'s -- both real and complex -- of $M_{2n}$. In (\ref{gen_func}), $\chi_{\mathbb{R}}(z) = 1$ if $z$ is real and 0 otherwise and $\langle \cdots \rangle_{M_{2n}}$ denotes an average over the joint distribution~(\ref{e:jpdf}), further summed over all possible $(l,m)$ \cite{BS09, Si07}. It turns out that such
averages (\ref{gen_func}) can be computed explicitly in terms of Pfaffians \cite{BS09, Si07}, as follows.   
Let $f\left(z\right)$ be any smooth integrable complex function, and $\left\{p_{j}\left(z\right)\right\}_{j=0}^{2n-1}$
be an arbitrary sequence of monic polynomials of degree $j$, then
	\begin{equation}\label{e:RMT_av}
		\left\langle \prod_{i=1}^{2n}
		f\left(\zeta_i\right)\right\rangle_{M_{2n}} = \frac{\Pf\left(U_f\right)}{\Pf\left(U_1\right)} \;,
	\end{equation}
where ${\rm Pf}$ denotes a Pfaffian \cite{footnote_pfaffian} and $U_f$ is a skew symmetric (i.e., anti-symmetric) matrix of size $2n\times 2n$ with entries
$u_{j,k} = (p_{j-1}f,p_{k-1}f)_{w}$ and skew product
	\begin{multline}\label{product}
		(h,g)_w = \int\limits_{\R^2} h\left(x\right)g\left(y\right)
		\sgn\left(y-x\right) w\left(x\right)w\left(y\right)\dx{x}\dx{y}
		\\+2i\int\limits_{\C}
		h\left(z\right)g\left(\bar{z}\right)
		\sgn\left[{\rm{Im}}(z)\right]
		w\left(z\right)w\left(\bar{z}\right)\dx{^2z} \;,
	\end{multline}
where $w(z)$ is given in (\ref{e:weight}). To compute the ratio in (\ref{e:RMT_av}), it is convenient to choose the monic polynomials $p_j(z)$ to be skew-orthogonal with respect to the product~(\ref{product}) (with this choice, the denominator in (\ref{e:RMT_av}) is easy to compute~\cite{SM}). 
Using these polynomials \cite{Fo10}, the GF in (\ref{gen_func}) can be evaluated explicitly using (\ref{e:RMT_av}), leading to 
%
%
%
%
%
\cite{GP18}
	\begin{equation}\label{e:Hilb_det}
	\langle e^{s\mathcal{N}_{n}} \rangle_{M_N}= \det\limits_{0\leq j,k\leq n-1}
	\left[\delta_{j,k} - \frac{1-e^{2s}}{\pi(j+k+1/2)}\right].
	\end{equation}
	Let us denote by ${H}_n$ the $n \times n$ matrix with entries $h_{j,k} = {(\pi(j+k+1/2))^{-1}}$. 
	We write the determinant in (\ref{e:Hilb_det}) as $\det({\mathbb{I}} - \alpha H_n) = \exp {\rm Tr}(\ln ({\mathbb{I}} - \alpha H_n))$, with $\alpha = (1-e^{2s})$
	and then expand the logarithm, to get $\det({\mathbb{I}} - \alpha H_n) = \exp[\sum_{m\geq 1} (\alpha^m/m) {\rm Tr}(H^m_n)]$. 
	The asymptotic analysis of the traces yields \cite{Wi66}	
	\begin{equation*}
	\Tr H^m_n = \frac{1}{2\pi}\int\limits_{0}^{\infty}
	\sech^m\left(\frac{\pi u}{2}\right)\dx{u}\log n\left(1+o(1)\right), n\to \infty.
	\end{equation*}
	By summing up these traces, we obtain 
	\begin{equation}\label{e:gen_fun}
		\langle e^{s\mathcal{N}_{n}}\rangle_{M_{2n}} = n^{\frac{1}{2\pi}
			\int\limits_{0}^{\infty}\log\left(1-(1-e^{2s})\sech\frac{\pi u}{2}\right)\dx{u}+o(1)}.
	\end{equation}
	For $s<0$, the integral can be calculated explicitly as 
	\begin{eqnarray}\label{e:psi}
	\langle e^{s\mathcal{N}_{n}}\rangle_{M_{2n}} \sim n^{\psi(s)} \,, \,  \psi(s) = \frac{1}{8} - \left[\frac{\sqrt{2}}{\pi} \cos^{-1} \left( \frac{e^{s}}{\sqrt{2}}\right)  \right]^{2} \;
	\end{eqnarray}
By taking $s\to -\infty$ we get the probability that $M_{2n}$
	has no real eigenvalues, using $Q_0(1,n) = {\rm Prob.}({\cal N}_n = 0) = \lim_{s \to -\infty} \langle e^{s\mathcal{N}_{n}}\rangle_{M_{2n}} \sim n^{-2 b}$. From Eq.~(\ref{e:psi}), we thus obtain $b = (-1/2) \lim_{s \to -\infty} \psi(s) = 3/16$, as announced in Eq. (\ref{exact_theta_2}). From the GF in (\ref{e:psi}), we also obtain the cumulants of ${\mathcal N}_n$. To export these results to the $2d$-diffusion equation, we recall that the number of zero-crossings $N_t$ identifies with the {\it positive} real eigenvalues $\mathcal{N}_{n}^+$ of $M_{2n}$. Hence, for $n \gg 1$, the number of positive and negative $\mathcal{N}_{n}^{\pm}$ real eigenvalues are both independent and identically distributed~\cite{GP18}, one obtains that $\langle e^{s\, {\cal N}_n^+} \rangle_{M_{2n}} \sim n^{\frac{\psi(s)}{2}}$. By further expanding $\psi(s)$ close to $s=0$ \cite{SM}, one finally obtains the result announced in Eq.~(\ref{cumulants}). Similarly, transposing this result $\langle e^{s\, {\cal N}_n^+} \rangle_{M_{2n}} \sim n^{\frac{\psi(s)}{2}}$ to the diffusion equation, one obtains the large deviation form in (\ref{large_dev}) with $\varphi(x) =\max_{s \in {\mathbb R}} [s x - \psi(s)/2]$~\cite{footnote_continuation}. From this relation, together with the expression for $\psi(s)$ in (\ref{e:psi}), we obtain the asymptotic behaviours given in Eq. (\ref{asym_phi})~\cite{SM}.

%
%
%

Several results found so far point out to an intriguing connection with the zero temperature Glauber dynamics of the 
Ising spin chain~\cite{DHP95, DHP96}. First, $b = 3/16$ is thus half of the persistence exponent, $\theta_{\rm Ising} = 3/8$, found there \cite{DHP95, DHP96}. In fact, $3/16$ is exactly the persistence exponent corresponding to the spin at the origin of the semi-infinite Ising chain \cite{DHP95,DHP96}. Furthermore, the expression found for $\psi(s)$ in Eq. (\ref{e:psi}) is strongly reminiscent of the expression found for the persistence exponent for the $q$-state Potts chain, with $T=0$ Glauber dynamics (see, e.g., Eq.~(2) of \cite{DHP95}). So what is this connection?

To understand it, let us come back to the $2d$-diffusion field $X(t) = {\phi({\bf 0},t)}/{\langle \phi({\bf 0},t)^2\rangle}$ and consider the ``clipped'' process $\sgn(X(t))$ \cite{footnote_clipped}. As recalled above $X(t)$ has the same statistical properties as the Kac's polynomials $K_n(x)$ in the limit $n \to \infty$ and $x \to 1$. Transposing recent results obtained for Kac's polynomials in the limit $n \to \infty$ \cite{MS13}, we can compute the multi-time correlation functions of $\sgn(X(t))$ which are given by Pfaffians~\cite{footnote_pfaffian} 
\begin{equation}\label{Pfaffian_sigma}
\langle  \sgn(X(t_1)) \cdots \sgn(X(t_{2m})) \rangle \sim {\rm Pf}(A)
\end{equation}
for $1\ll t_1\ll t_2 \ll \cdots \ll t_{2m}$, and where $A = (a_{i,j})_{1 \leq i,j \leq 2m}$ is a $2m \times 2m$ anti-symmetric matrix with $a_{i,i} = 0$ and for $i < j$, $a_{i,j} = - a_{j,i}$ where
\begin{eqnarray}\label{2point}
a_{i,j} =  \frac{2}{\pi}  \sin^{-1} \left( \langle X(t_i) X(t_j)\rangle \right) = \frac{2}{\pi} \sin^{-1} \left( \frac{2 \sqrt{t_i\, t_j} } {t_i + t_j}\right)\;.
\end{eqnarray}
By symmetry, the even correlation functions vanish. For $m=1$, Eqs.~(\ref{Pfaffian_sigma}) and (\ref{2point}) hold for any normalised Gaussian process. However, for $m > 1$, this Pfaffian structure, which holds for the GSP $Y(T) = X(e^{T})$, is~nontrivial. 

Let us now consider the semi-infinite Ising spin chain, whose configuration at time $t$ is given by $\{\sigma_i(t)\}_{i \geq 0}$, with $\sigma_i(t) = \pm 1$. Initially, $\sigma_i(0) = \pm 1$ with equal probability $1/2$ and, at subsequent time, the system evolves according to the Glauber dynamics at $T=0$ (see \cite{DHP95,DHP96} for details). 
%
%
Using the formulation of the dynamics in terms of coalescing random walks \cite{DHP95,DHP96}, we show that the multi-time correlation functions of $\sigma_0$ are also given by the same Pfaffian formula (\ref{Pfaffian_sigma}), namely, for $1 \ll t_1 \ll t_2 \ll \cdots \ll t_{2m}$ \cite{SM}
\begin{eqnarray}\label{Pfaffian_sigma0}
\langle \sigma_0(t_1) \cdots \sigma_0(t_{2m})\rangle \sim {\rm Pf}(A) \;, 
\end{eqnarray} 
with precisely the same anti-symmetric matrix $A$ (\ref{2point}). Therefore, we conclude that $\sgn(X(t))$ for the $2d$-diffusion equation and $\sigma_0(t)$ in the semi-infinite Ising chain with Glauber dynamics are actually the same process in the large time limit \cite{foot_smooth}. One can then use the known result for the persistence of $\sigma_0(t)$ \cite{DHP95,DHP96} to conclude that $b=3/16$, as found above by a completely different method. Note that the exact relation found here $\sigma_0(t) \propto {\rm sgn}(\phi({\bf 0},t))$ (for $t \gg 1$), where $\phi({\bf x},t)$ is the $2d$-diffusing field, is reminiscent, albeit different from, the so-called OJK approximate theory \cite{OJK82} in phase ordering kinetics \cite{Bra02}, which instead approximates the $1d$ spin field by the sign of the $1d$ diffusing field.    
	
	To conclude, we have computed exactly the persistence exponent of $2d$-diffusion equation, or equivalently the one of Kac's polynomials, $\theta(2) = b =3/16$. This was done in two different ways: (i) by using the connection to truncated random orthogonal matrices, and for which our results are actually mathematically rigorous \cite{GP18}, (ii) by establishing an exact mapping to the semi-infinite Ising chain with Glauber dynamics at $T=0$ (see Fig. \ref{Fig:recap}). Thanks to (i), we computed the full statistics of the number of the zero crossings (\ref{large_dev})-(\ref{cumulants}). These RMT tools will certainly be useful to compute other properties of the GSP with correlator $c(T) = {\rm sech}(T/2)$ and of the different physical models associated to it (see Fig. \ref{Fig:recap}). 
	
	\begin{acknowledgements}
	G. S. wishes to thank warmly S. N. Majumdar for illuminating discussions and ongoing collaborations on this topic and related ones. We would also like to acknowledge A. Dembo, I. Dornic, Z. Kabluchko, M. Krishnapur and P. Le Doussal for useful discussions and comments. We also thank Les Houches school of physics, where this project was initiated. This work was partially supported by EPSRC EP/N009436/1 as well as by the ANR grant ANR-17-CE30-0027-01 RaMaTraF. 
	\end{acknowledgements}

%

	\onecolumngrid
	
	\newpage
	
	
	\begin{large}
	\begin{center}
		{\bf Supplementary Material for {\it Exact persistence exponent for the $2d$-diffusion 
			equation and related Kac polynomials}}
			\end{center}
	\end{large}
	
	We give the principal details of the calculations described in the main text of the Letter.

	\section{1) Derivation of Eqs. (9) and (10)  in the main text}
	In this section we discuss the ideas behind the formulae \eqref{identity_det} and \eqref{series_f}
	in order to give a clear picture of the connection between truncations of
	random orthogonal matrices and Kac polynomials. Let $O$ be a $\left(2n+1\right)\times \left(2n+1\right)$ orthogonal matrix decomposed as [see Eq. (8) in the main text]
	\begin{equation}\label{def_O}
	O = 
		\begin{pmatrix}
			M_{2n} & \mathbf{u} \\
			\mathbf{v}^T & a
		\end{pmatrix},
	\end{equation}
	where $\mathbf{u}$ and $\mathbf{v}$ are column vectors of length $Žn$ 
	and $a$ is a scalar. We are interested in eigenvalues of $M_{2n}$ that
	can be found as roots of the characteristic polynomial 
	$g_M\left(z\right) = \det\left(z{\mathbb I}_{2n}-M_{2n}\right)$. 
	
	We first recall that, for a block matrix $U$ of the form
	\begin{eqnarray}\label{def_block}
	U = 
	\begin{pmatrix}
	A & B \\
	C & D
	\end{pmatrix} \;,
	\end{eqnarray}
	where $A, B, C$ and $D$ are matrices (with $A$ and $D$ invertible), the determinant $\det U$ can be written in two different ways as
	\begin{eqnarray}
	\det U &=& \det A \, \det (D- CA^{-1}B) \label{det_block1} \\
	&=& \det D \, \det(A - B D^{-1} C) \label{det_block2} \;.
	\end{eqnarray}
	Let us consider the following $(2n+1)\times(2n+1)$ rectangular matrix $X$~\cite{Kr09}
	\begin{equation*}
		X = 
		\begin{pmatrix}
			{\mathbb I}_{2n}-zM_{2n} & \mathbf{u} \\
			-\mathbf{v}^T & z^{-1}a
		\end{pmatrix},
	\end{equation*}
	then for $a\neq 0$ and assuming that ${\mathbb I}_{2n}-z\,M_{2n}$ is an invertible matrix, one can write (using Eqs. (\ref{det_block1}) and (\ref{det_block2}))
	\begin{eqnarray}
		\det X &=& \det \left({\mathbb I}_{2n}-zM_{2n}\right)\left(z^{-1}a+
		\mathbf{v}^T\left({\mathbb I}_{2n}-zM_{2n}\right)^{-1}\mathbf{u}\right) \label{det1.1}\\
		&=& z^{-1}a \, \det\left({\mathbb I}_{2n}-zM_{2n}+za^{-1}\mathbf{u}\mathbf{v}^T\right) \;. \label{det1.2}
	\end{eqnarray}
	In Eq. (\ref{det1.2}) one recognizes the inverse of the the top left corner of $O^{-1}$. Indeed, the block inversion formula gives 
	\bea\label{inverse_formula}
	O^{-1} =
	\begin{pmatrix}
	(M_{2n}- a^{-1}\, {\bf u} {\bf v}^T)^{-1} & \tilde B \\
	\tilde C & \tilde D
	\end{pmatrix}
	\eea
	where the matrices $\tilde B, \tilde C$ and $\tilde D$ are not needed here. Since $O$ is an orthogonal matrix, $O^{-1} = O^T$ and hence (from Eqs. (\ref{def_O}) and (\ref{inverse_formula})) we obtain
	\bea\label{relation}
	   (M_{2n}- a^{-1}\, {\bf u} {\bf v}^T)^{-1} = M_{2n}^{T} \Longrightarrow M_{2n}- a^{-1}\, {\bf u} {\bf v}^T = (M_{2n}^{-1})^T
	\eea
	Therefore, Eq. (\ref{det1.2}) can be written as 
	\begin{eqnarray}
		\det X = z^{-1}a\det\left({\mathbb I}_{2n} - z(M_{2n}^{-1})^T\right) = z^{-1}a \det M_{2n}^{-1}\det\left(M_{2n}-z{\mathbb I}_{2n}\right).
	\end{eqnarray}
	This yields 
	\begin{equation}\label{main_identity}
		\frac{\det\left(zI_{2n}-M_{2n}\right)}{\det\left(zM_{2n}-I_{2n}\right)} =  
		\det M_{2n} \left(1+za^{-1}\mathbf{v}^T\left(I_{2n}-zM_{2n}\right)^{-1}\mathbf{u}\right)
		= \frac{\det M_{2n}}{a} \left(a+z\mathbf{v}^T\left(I_{2n}-zM_{2n}\right)^{-1}\mathbf{u}\right).
	\end{equation}
	Notice that the inverse of the block matrix $O$ in (\ref{def_O}) can also be written as
	\begin{eqnarray}\label{inverse2}
	O^{-1} = 
	\begin{pmatrix}
	\tilde A' & \tilde B' \\
	\tilde C' & (a - {\bf v}^T M_{2n}^{-1} {\bf u})^{-1} 
	\end{pmatrix} \;,
	\end{eqnarray}
       	where the matrices $\tilde A', \tilde B'$ and $\tilde C'$ are not needed here. Since $O$ is an orthogonal matrix, $O^{-1} = O^T$ and hence (from Eqs. (\ref{def_O}) and (\ref{inverse2})) we obtain
	\begin{eqnarray}\label{identity2}
	a = (a - {\bf v}^T M_{2n}^{-1} {\bf u})^{-1} \;.
	\end{eqnarray}
	On the other hand, applying the formula in Eq. (\ref{det_block1}) to the block matrix $O$ in (\ref{def_O}), we obtain
	\begin{eqnarray}\label{detO}
	\det O = \det M_{2n} \det(a - {\bf v}^T M_{2n}^{-1} {\bf u}) = \det{O} = \frac{\det M_{2n}}{a} \;,
	\end{eqnarray}
	where, in the last equality, we have used the identity (\ref{identity2}). Finally, by combining Eq. (\ref{main_identity}) and (\ref{detO}), we obtain the formula given in Eq. (9) in the main text.

	With probability one the eigenvalues of $M_{2n}$ are interior points of the unit disc, since
	$M_{2n} \, M_{2n}^T = \mathbb{I}-\mathbf{u}\mathbf{u}^T < \mathbb{I}$. For $\left|z\right|<1$
	one can write the r.h.s. as a series (up to the sign of $\det O$)
	\begin{equation}
		F_{2n}\left(z\right) = a+\sum\limits_{k=1}^{\infty}z^k \mathbf{v}^TM_{2n}^{k-1}\mathbf{u} \;.
	\end{equation}
	All the roots of the series which are inside of the unit circle are the eigenvalues of matrix
	$M_{2n}$ and vice versa. Finally, one can show \cite{HKPV09} that the scaled sequence of 
	the coefficients of the series $F_{2n}$ given by $\sqrt{2n}\{a, \mathbf{v}^T\,\mathbf{u}, 
	\mathbf{v}^T\,M_{2n}\,\mathbf{u}, \mathbf{v}^TM_{2n}^{2}\,\mathbf{u}, \cdots\}$ converges, 
	as $n \to \infty$,  to the sequence of i.i.d. Gaussian random variables, with zero mean and unit variance.
	This implies that, for large $N$, the eigenvalues of truncated random orthogonal matrices behave
	identically to zeros of random Kac series, studied in \cite{MS13} and discussed above. {And one can check that,
	if we put $n \to \infty$ in the kernel of the Pfaffian point process describing 
	the distribution of eigenvalues of random matrices $M_{2n}$, then the corresponding limiting kernel
	coincides with the one obtained in \cite{MS13} (see, \cite{Fo10} for details)}.

	\section{2) Derivation of the formula given in Eq. (17) in the main text}

	We present the main steps leading to Eq. (17) in the main text and refer reader to \cite{GP18}
	for further details. Let $M_{2n}$ be a truncated random orthogonal matrix, as defined in Eq. (\ref{def_O}).	
	Its spectrum consists of $\ell$ real eigenvalues $\lambda_1\leq\ldots\leq\lambda_{\ell}$ and
	$m = n-\ell/2$ pairs of complex conjugate eigenvalues $z_{1,2}=x_1 \pm iy_1, \ldots, 
	z_{2m-1,2m} = x_m\pm iy_m$ whose joint distribution was derived in \cite{KSZ10} and given by 
	\eqref{e:jpdf}. It is well-known in random matrix theory (see, eg \cite{Si07,BS09}) that such
	conditional distributions give rise to Pfaffian point processes which satisfy \eqref{e:RMT_av}
	\cite{Si07}. Choosing the polynomials
	\begin{equation}\label{ortho_p_1}
		P_{2j}\left(z\right) = z^{2j}, P_{2j+1}\left(z\right) = 
		z^{2j+1}-\frac{2j}{2j+1}z^{2j-1} \;,
	\end{equation}
	which are orthogonal with respect to the skew product in Eq. (16) in the main text~ \cite{Fo10}, we get 
	\begin{equation*}			
		\left\langle \prod_{\zeta \in \mathrm{spec}M_{2n}}
		f\left(\zeta\right)\right\rangle_{M_{2n}} = \Pf\left(U_f\right)/\Pf\left(U_1\right),
	\end{equation*}
	where
	\begin{eqnarray*}
		\left(U_f\right)_{i,j} &=& \int\limits_{\R^2} 
		\sgn\left(y-x\right) f\left(x\right)
		f\left(y\right) P_{i-1}\left(x\right)P_{j-1}\left(y\right)
		w\left(x\right)w\left(y\right)\dx{x}\dx{y}
		\\&&+2i\int\limits_{\C}
		f\left(z\right)f\left(\bar{z}\right)
		P_{i-1}\left(z\right)P_{j-1}\left(\bar{z}\right)
		\sgn\left(\Im z\right)
		w\left(z\right)w\left(\bar{z}\right)\dx{^2z}, \\
		\left(U_1\right)_{i,j} &=& \int\limits_{\R^2} 
		\sgn\left(y-x\right) P_{i-1}\left(x\right)P_{j-1}\left(y\right)
		w\left(x\right)w\left(y\right)\dx{x}\dx{y}
		+2i\int\limits_{\C}
		P_{i-1}\left(z\right)P_{j-1}\left(\bar{z}\right)
		\sgn\left(\Im z\right)
		w\left(z\right)w\left(\bar{z}\right)\dx{^2z}.
	\end{eqnarray*}
	Let us denote by $\left(U^{(r)}\right)_{i,j}, \left(U^{(c)}\right)_{i,j}$ the two integrals with respect
	to real and complex variables in the right hand side (r.h.s.) of the above expressions. Then one can see that with 
	$f(z) = 1-(1-e^s)\chi_{\R}(z) $ we have
	\begin{equation}\label{relation_U}
		\left(U_f\right)_{i,j} = e^{2s}\left(U^{(r)}\right)_{i,j} + \left(U^{(c)}\right)_{i,j}
		= \left(U_1\right)_{i,j} + (e^{2s}-1)\left(U^{(r)}\right)_{i,j}.
	\end{equation}
	We now focus on the computation of both $\left(U^{(r)}\right)_{i,j}$ and $\left(U^{(c)}\right)_{i,j}$ for $i,j = {1,2, \cdots,2n}$.
	First we note that both integrals vanish if $i$ and $j$ are of the same parity. 
	This can be seen from
	\begin{eqnarray*}
	\left(U^{(r)}\right)_{i,j} &= &
	\frac{1}{2\pi}
	\iint\limits_{-1}^{1} \sgn\left(y-x\right) P_{i-1}(x)P_{j-1}(y)
	\left(1-x^2\right)^{-\frac{1}{2}}\left(1-y^2\right)^{-\frac{1}{2}}
	\dx{x}\dx{y}
	\\	
	\Bigg|x,y\to -x,-y\Bigg|
	&=& 
	\frac{1}{2\pi}(-1)^{i+j}
	\iint\limits_{-1}^{1} \sgn\left(x-y\right) P_{i-1}(x)P_{j-1}(y)
	\left(1-x^2\right)^{-\frac{1}{2}}\left(1-y^2\right)^{-\frac{1}{2}}
	\dx{x}\dx{y} = -\left(U^{(r)}\right)_{i,j}.
	\end{eqnarray*}
	For complex part we write
	\begin{eqnarray*}
	\left(U^{(c)}\right)_{i,j} &= &
	\frac{i}{\pi}\iint\limits_{\D} \sgn\left(y\right)
	P_{i-1}\left(x+iy\right)P_{j-1}\left(x-iy\right)	
	\frac{\dx{x}\dx{y}}{\left((1-x^2+y^2)^2+4x^2y^2\right)^{1/2}}.
	\end{eqnarray*}
	After expanding both polynomials in powers of $x,y$ we can see that the terms with odd powers
	of $x$ and even powers of $y$ cancel due to the symmetry of integrand. If $i$ and $j$ have 
	the same parity, then there are no other terms in the expansion. Besides, it is easy to see
	that both $\left(U^{(r,c)}\right)_{i,j}$ are skew symmetric by interchanging $i \leftrightarrow j$
	and we study below only the terms of the form $\left(U^{(r,c)}\right)_{2p+1,2q+2}$. For the real part we obtain
	\begin{eqnarray}
		\left(U^{(r)}\right)_{2p+1,2q+2} &= &
		\frac{1}{2\pi}
		\iint\limits_{-1}^{1} \sgn\left(y-x\right) x^{2p} \left(y^{2q+1} - \frac{2q}{2q+1}y^{2q-1}\right)
		\frac{\dx{x}\dx{y}}{\sqrt{1-x^2}\sqrt{1-y^2}} \nonumber
		\\
		&=&
		\frac{1}{2\pi (2q+1)}
		\int\limits_{-1}^{1}x^{2p}\frac{\dx{x}}{\sqrt{1-x^2}}\int\limits_{-1}^{1}\sgn(y-x) 
		\frac{\dx{}}{\dx{y}} \left(-y^{2q}\sqrt{1-y^2}\right)
		\dx{y}
		\nonumber \\
		&=&
		\frac{1}{\pi (2q+1)}
		\int\limits_{-1}^{1}x^{2p+2q}\dx{x} = \frac{2}{\pi (2q+1)(2p+2q+1)} \;. \label{real}
	\end{eqnarray}
	For the complex part we write
	\begin{eqnarray}\label{U_inter}
		\left(U^{(c)}\right)_{2p+1,2q+2} &= &
		\frac{2i}{\pi}
		\iint\limits_{\D_+}
		\left(x+iy\right)^{2p} \left(\left(x-iy\right)^{2q+1} - \frac{2q}{2q+1}\left(x-iy\right)^{2q-1}\right)
		\frac{\dx{x}\dx{y}}{\left((1-x^2+y^2)^2+4x^2y^2\right)^{1/2}},
	\end{eqnarray}	
	and use
	\begin{equation*}
		\left(
		\frac{\partial}{\partial x} + i \frac{\partial}{\partial y}
		\right)
		\left(x-iy\right)^{2q} \frac{\sqrt{\left(1-x^2+y^2\right)^2+4x^2y^2}}{1-\left(x+iy\right)^2}
		= -2(2q+1)		
		\frac{\left(\left(x-iy\right)^{2q+1} -
		 \frac{2q}{2q+1}\left(x-iy\right)^{2q-1}\right)}{\left((1-x^2+y^2)^2+4x^2y^2\right)^{1/2}} \;.
	\end{equation*}
 Performing the integration in (\ref{U_inter}) we get
	\begin{eqnarray}
		\left(U^{(c)}\right)_{2p+1,2q+2} &=& -\frac{i}{\pi(2q+1)}
		\left[
		\int\limits_{0}^{1} \dx{y} 
		(x+iy)^{2p}(x-iy)^{2q}
		\frac{\sqrt{\left(1-x^2+y^2\right)^2+4x^2y^2}}{1-\left(x+iy\right)^2} 
		\Bigr|_{x=-\sqrt{1-y^2}}^{x=\sqrt{1-y^2}}
		\right.
		\nonumber \\
		&&
		\left.
		+i\int\limits_{-1}^{1}\dx{x}
		(x+iy)^{2p}(x-iy)^{2q}
		\frac{\sqrt{\left(1-x^2+y^2\right)^2+4x^2y^2}}{1-\left(x+iy\right)^2} 
		 \Bigr|_{y=0}^{y=\sqrt{1-x^2}}
		\right]
		\nonumber \\
		&=&
		\frac{1}{\pi(2q+1)}	\left(1+(-1)^{p-q}\right)
		\int\limits_{-1}^{1} \left(\sqrt{1-x^2}+ix\right)^{2p-1}
		\left(\sqrt{1-x^2}-ix\right)^{2q}\dx{x}
		-\frac{1}{\pi(2q+1)}\int\limits_{-1}^{1} x^{2p+2q}\dx{x} \nonumber \\
		&=& \frac{1}{(2q+1)} \delta_{p,q} 
		-\frac{2}{\pi(2q+1)(2p+2q+1)} \;. \label{complex}
	\end{eqnarray}
	By adding the real (\ref{real}) and complex (\ref{complex}) parts together we obtain [using Eq. (\ref{relation_U})] $\left(U_1\right)_{i,j} = \frac{1}{j-1},$ 
	if $i=j-1$ is odd, $\left(U_1\right)_{i,j} = - \frac{1}{j},$ if $i=j+1$ is even
	and zero otherwise. One can then easily check that
	\begin{equation*}
		\Pf U_1 = \prod\limits_{j=1}^n (2j-1)^{-1},
	\end{equation*}
	and use the above formulae in Eqs. (\ref{real}) and (\ref{complex}), together with Eq. (\ref{relation_U}),  to show the formula (17) given in the text.  
	
\section{3) Derivation of the formula for the cumulants given in Eq.~(\ref{cumulants})}
	
	In this section, we derive the expression for the cumulants $\kappa_p$ of the number of zero crossings $N_t$ of the $2d$-diffusing field up to time $t$, given in Eq. (7) in the main text. As explained in the text, the statistical properties of $N_t$ for large $t$ are obtained from the number of positive real eigenvalues ${\cal N}^+_n$ of the truncated orthogonal random matrix $M_{2n}$ [see Eq. (\ref{def_O})] for which we have shown in the paper that
	\begin{eqnarray}\label{start_cumul}
	\langle e^{s\,{\cal N}_n^+} \rangle_{M_{2n}} \sim n^{\frac{\psi(s)}{2}}
	\end{eqnarray}
	where $\psi(s)$ is given in Eq. (19) in the main text. The cumulants of ${\cal N}^+_n$, denoted as $\langle [{\cal N}^+_n]^p\rangle^c_{M_{2n}}$ can be obtained from the following expansion 
	\begin{eqnarray}\label{series_cumul}
	\ln \langle e^{s\,{\cal N}_n^+} \rangle_{M_{2n}}  = \sum_{p=1}^\infty \frac{s^p}{p!} \langle [{\cal N}^+_n]^p\rangle^c_{M_{2n}} \sim \frac{\psi(s)}{2} \ln n \;,
	\end{eqnarray}
	where in the second estimation we have used (\ref{start_cumul}). Therefore, to compute these cumulants, we need to expand $\psi(s)$ for small $s$. It is then convenient to start from Eq. (18) and expand the logarithm to get
	\begin{equation}\label{smalls_1}
	\psi(s) = -\frac{1}{2 \pi} \sum_{m=1}^\infty \frac{1}{m}(1-e^{2s})^m\int_0^\infty du \, {\rm sech}^m\left( \frac{\pi u}{2}\right) = - \frac{1}{2 \pi^2} \sum_{m=1}^\infty \frac{1}{m} 
	B\left( \frac{m}{2},\frac{1}{2}\right) (1 - e^{2 s})^m \;,
	\end{equation}
	valid for $s\leq 0$ where we have used
	\begin{equation}\label{integral_sech}
	\int_0^\infty du \, {\rm sech}^m\left( \frac{\pi u}{2}\right) = 
	\frac{1}{\pi} B\left( \frac{m}{2},\frac{1}{2}\right) \;,
	\end{equation}
	with $B(a,b)$ being the Euler beta function. We now expand $(e^{2s} - 1)^m$ in (\ref{smalls_1}), which yields
	\begin{equation}\label{smalls_2}
		(e^{2s}-1)^m = \left(\sum\limits_{p=1}^{\infty} \frac{(2s)^p}{p!}\right)^m
		= \sum\limits_{p=m}^{\infty}\frac{(2s)^p}{p!} \sum\limits_{p_1+\ldots+p_m = p, p_i\geq 1}
		\frac{p!}{p_1!p_2!\ldots p_m!}.
	\end{equation}
	{Let us denote the last sum by $c_{m,p}$.} {We first notice that the binomial theorem yields
	\begin{equation*}
		\sum\limits_{p_1+\ldots+p_m = p, p_i\geq 0}
		\frac{p!}{p_1!p_2!\ldots p_m!} = \left(1+1+\ldots+1\right)^p = m^p.
	\end{equation*}
	However, to calculate $c_{m,p}$ we need to subtract all terms with $p_i = 0$ for at least one $1\leq i \leq m$.
	For this purpose, we use the inclusion-exclusion principle as follows. First we subtract all the terms with $p_j=0$ for a given
	value of $j$. There are $m$ choices of index $j$ and all terms with $p_j=0$ sum up to
	\begin{equation*}
	\sum\limits_{\substack{p_1+\ldots+p_m = p\\  p_j=0, \forall i\neq j \,\, p_i \geq 0}}
	\frac{p!}{p_1!p_2!\ldots p_m!} = \left(m-1\right)^p.
	\end{equation*}
	Next we add all the terms containing $p_j=p_k=0$ for a given pair $j<k$. These terms sum up to
	\begin{equation*}
	\sum\limits_{\substack{p_1+\ldots+p_m = p\\  p_j=p_k=0, \forall i\neq j,k \,\, p_i \geq 0}}
	\frac{p!}{p_1!p_2!\ldots p_m!} = \left(m-2\right)^p,
	\end{equation*}
	and there are $\binom{m}{2}$ choices of pair of indices $j<k$. Continuing the calculation we obtain} 
	
	\begin{equation}\label{cmp}
		c_{m,p} = m^p - m(m-1)^p + \binom{m}{2}(m-2)^p-\ldots = \sum\limits_{\ell=0}^{m} (-1)^{\ell} 
		\binom{m}{\ell}(m-\ell)^p = m! \mathcal{S}^{(m)}_p,
	\end{equation}
	where, in the last equality, we have used the definition of Stirling number of the second kind $\mathcal{S}^{(m)}_p$ \cite{Stirling_supp}. 
	Combining Eqs. (\ref{smalls_1}), (\ref{smalls_2}) and (\ref{cmp}) we find
	\begin{equation}\label{smalls_3}
		\psi(s) = -\frac{1}{2\pi^2}\sum\limits_{m=1}^{\infty}
		(-1)^m (m-1)!B\left(\frac{m}{2},\frac{1}{2}\right)
		\sum\limits_{p=m}^{\infty} \frac{(2s)^p}{p!}\mathcal{S}^{(m)}_p.
	\end{equation}
	One can the use induction and the recurrence relations for Stirling numbers to show
	\begin{equation}
		\mathcal{S}_p^{(m)} \leq 2^p \frac{p!}{m!},
	\end{equation}
	which, together with the asymptotic behaviour for $B\left(\frac{m}{2},\frac{1}{2}\right) \sim 
	\sqrt{\frac{2\pi}{m}}$ for $m\gg1$, yields that the double infinite sum in (\ref{smalls_3}) is absolutely convergent for small 
	values of $s$. One can thus interchange the order of the summation over $m$ and $p$ to obtain
	\begin{equation}\label{final_psi}
		\psi(s) = \sum\limits_{p=1}^{\infty} s^p \frac{2^{p-1}}{\pi^2 p!}
		\sum\limits_{m=1}^p (-2)^{m-1}\mathcal{S}_p^{(m)} \Gamma^2\left(\frac{m}{2}\right) \;.
	\end{equation}
Finally, by combining Eqs. (\ref{series_cumul}) together with the small $s$ expansion of $\psi(s)$ in (\ref{final_psi}), we obtain that $\langle [{\cal N}^+_n]^p\rangle^c_{M_{2n}} \sim \kappa_p \, \ln n $ where the expression for the cumulants $\kappa_p$ is given in Eq. (7) in the main text.

   \section{4) Asymptotic expansion of the large deviation function $\varphi(x)$}

In this section, we study the large deviation function $\varphi(x)$	defined in Eq. (5) and obtain its asymptotic behaviours given in (6). This function $\varphi(x)$ is defined in term from $\psi(s)$ given in Eq. (9) of the text
\begin{eqnarray}\label{def_phi_supp}
\varphi(x) = \max_{s \in \mathbb{R}} \left(s\,x - \frac{\psi(s)}{2}\right) \;,Ê\; \psi(s) = \frac{1}{8} - \frac{2}{\pi^2} \left(\cos^{-1} \left( \frac{e^s}{2}\right) \right)^2 \;.
\end{eqnarray}
Note that the function $\psi(s)$, initially defined for $s<0$ (see Eq. (18) in the text)	can be analytically continued to any positive $s>0$, where $\psi(s)$ is actually also real. The maximum of $(s\,x - \frac{\psi(s)}{2})$ in Eq. (\ref{def_phi_supp}) is attained at a unique point $s^*$ which satisfies
\begin{eqnarray}\label{derivative}
x - \frac{1}{2} \psi'(s^*) = 0 \Longleftrightarrow x  = {\cal F}(z^*) \; {\rm with} \;  
\begin{cases}
&\;\;\;\;z^*= \dfrac{e^{s^*}}{2} \\
&{\cal F}(z) = \dfrac{2\,z}{\pi^2 \sqrt{1-z^2}} \cos^{-1}(z) \;.
\end{cases}
\end{eqnarray}
Note that the function ${\cal F}(z)$ is a monotonously increasing function such that ${\cal F}(z \to 0) \to 0$ and ${\cal F}(z \to \infty) \to \infty$ such that its inverse ${\cal F}^{-1}(z)$ is well defined for any $z \in {\mathbb R}^+$. In particular, one can show that its asymptotic behaviours are given by
\begin{eqnarray}\label{asympt_F-1}
{\cal F}^{-1}(z) \sim
\begin{cases}
&\pi z + 2 \pi z^2 \;, z \to 0\\
&\frac{1}{2}\, e^{\frac{\pi^2}{2}\,z} \;, \; \hspace*{0.4cm} z \to \infty \;.
\end{cases}
\end{eqnarray}
Therefore using Eqs. (\ref{def_phi_supp}) and (\ref{derivative}), we obtain that $\varphi(x)$ reads
\bea\label{explicit_phi} 
\varphi(x) = x\, \ln(\sqrt{2}\, z^*) - \frac{1}{16} - \frac{\pi^2}{4}\,x^2  + \frac{\pi^2}{4}\left(\frac{x}{z^*} \right)^2  \;,\; z^* = {\cal F}^{-1}(x) \;.
\eea
Using this expression (\ref{explicit_phi}) together with the asymptotic behaviours of ${\cal F}^{-1}(x)$ in Eq. (\ref{asympt_F-1}), we obtain the asymptotic behaviours of $\varphi(x)$ as
\begin{eqnarray}\label{asympt_phi_supp}
\varphi(x) \sim
\begin{cases}
&\dfrac{3}{16} + x\ln x + x(\ln(\pi \sqrt{2})-1) \;, \; x \to 0 \\
&\\
&\dfrac{\pi^2}{4}\,x^2 - \dfrac{x}{2} \ln 2 - \dfrac{1}{16} \;, \; x \to \infty
\end{cases}
\end{eqnarray}
the first terms of which give the behaviours given in Eq. (6) in the text.

We end this section by a remark: here we have studied the distribution $\tilde p_k(n)$, with $k=0, \cdots, 2n$, of the number ${\cal N}_n^+$ of the real (positive) eigenvalues of $M_{2n}$ in the limit where both $k$  and $n$ are large, and $k = O(\ln n)$, and this regime $\tilde p_k(n) \sim n^{-\varphi(k/\ln n)}$. However, there might well exist different scaling regimes, for instance for $k = O(n)$, as in the case of the real Ginibre matrices \cite{real_ginibre_supp}. In fact, for truncated orthogonal matrices the case $k=2n$ was investigated in \cite{FK17_supp} and it would be interesting to study the generic case $\alpha = k/(2n)$, for $k$ and $n$ large.

	\section{5) Real random series as a Pfaffian Point Process and the formulae (20)-(21)}
	
	In this section, we recall the main results obtained in Ref. \cite{MS13_supp} and derive the expression given in Eqs. (20) and~(21) in the text. In Ref. \cite{MS13_supp}
	where the 
	authors considered the "limiting case" of Kac polynomials of degree $n \to \infty$ and studied the distribution
	of the associated roots. More precisely, let $\left\{a_k\right\}_{k=0}^{\infty}$ be a sequence of i.i.d.
	real standard Gaussian random variables and let
	\begin{equation*}
		f\left(z\right) = \sum\limits_{k=0}^{\infty} a_k z^k,
	\end{equation*}
	be an analytic function inside the unit disk $\mathbb{D}$. As a function of the real parameter $x \in ]-1,1[$, $\left\{f(x)\right\}_{-1<x<1}$
	is a real Gaussian process 
	with covariance kernel $R(x,y) = \E[f(x) f(y)] = (1-x\,y)^{-1}$. 
	The real zeros of $f$ are random points on $\left(-1,1\right)$ and their distribution can be studied 
	by using the remarkable formula due to Hammersley \cite{Ham_supp} for the $m$-point correlation function $\rho_m(x_1, \cdots, x_m)$ of real zeroes. It reads
	\begin{equation}\label{e:MS_dens}
		\rho_m\left(x_1,x_2,\ldots,x_m\right) = 
		\frac{\E\left[|f'(x_1)f'(x_2)\ldots f'(x_m)|\rvert f(x_1)=f(x_2)=\ldots=f(x_m)=0\right]}{
			(2\pi)^{m/2}\sqrt{\det(R(x_i,x_j))}},
	\end{equation}
	The r.h.s. can be also rewritten by noting that
	\begin{multline}\label{Pf1}
		\left(\frac{2}{\pi}\right)^{m/2}\frac{1}{\sqrt{\det R(x_i,x_j)}}
		\E\left[|f'(x_1)f'(x_2)\ldots f'(x_m)|\rvert f(x_1)=f(x_2)=\ldots=f(x_m)=0\right]\\
		=		
		\lim\limits_{s_1\to x_1+0}\lim\limits_{s_2\to x_2+0}\ldots\lim\limits_{s_n\to x_m+0}
		\frac{\partial^m}{\partial x_1\partial x_2\ldots\partial x_m}
		\E\left[\sgn f(x_1) \sgn f(x_2)\ldots \sgn f(x_m)
		\sgn f(s_1)\sgn f(s_2)\ldots \sgn f(s_m)\right].
	\end{multline}
	The above formula clearly shows the importance of the so-called sign-correlation function for the 
	Gaussian process $f(t)$ defined as
	\begin{equation}\label{def_S}
		S\left(x_1,x_2,\ldots,x_{2k}\right) = 
		\E\left[\sgn f(x_1)\sgn f(x_2)\ldots \sgn f(x_{2k})\right].
	\end{equation}	

	The computation of the r.h.s. of \eqref{e:MS_dens} can then be carried out as follows. Since $f(.)$ is a Gaussian process of covariance $R(x,y)$, 
	one can show that the process $f(.)$ {\it conditioned} on $f(x_0)=0$ for some $x_0 \in ]-1,1[$, denoted as $(f(.)|f(x_0)=0)$ is also a Gaussian process. {Let $x_1, x_2,\ldots, x_m \in ]-1,1[$, then the Gaussianity 
	of $f(.)$ yields that the vector $\vec{f_0} = (f(x_0), f(x_1),\ldots,f(x_m))$ is normally distributed
	with zero mean and covariance matrix $\Sigma_0 = \left\{R(x_i,x_j)\right\}_{i,j=0}^{m}$.
	A simple calculation yields the distribution of the vector $\vec{f} = \left(f(x_1),\ldots,f(x_m)\right)$ conditioned on the event $f(x_0)=0$ as
	\begin{equation*}
		p\left(\vec{f}\right) \propto 
		\exp\left\{
			-\frac{1}{2}\vec{f}\,
			\Sigma_{22}\, \vec{f}^T\right\},
			\quad \mbox{where} \quad \Sigma_0^{-1} = 
			\begin{pmatrix}
				\Sigma_{11} & \Sigma_{12} \\
				\Sigma_{12}^T & \Sigma_{22}
			\end{pmatrix},
	\end{equation*}
	where $\Sigma_{11}$ is a scalar, $\Sigma_{12}$ is a row vector of size $m$ and $\Sigma_{22}$
	is an $m\times m$ matrix. The block inversion formula yields 
	\begin{equation*}
		\left(\Sigma_{22}\right)_{ij} = R\left(x_i,x_j\right) - R(x_i,x_0)R^{-1}(x_0,x_0)R(x_0,x_j).
	\end{equation*}
	This proves that the conditional process $(f(.)|f(x_0)=0)$ is Gaussian}	
	 with covariance
	\bea \label{Rtilde}
	\tilde R(x,y) = R(x,y) - \frac{R(x,x_0) R(x_0,y)}{R(x_0,x_0)} \;.
	\eea 
	For the special case considered here where $R(x,y) = 1/(1-xy)$, it is easy to check that 
	\begin{equation}\label{identity}
		\tilde R(x,y)=R(x,y) - \frac{R(x,x_0)R(y,x_0)}{R(x_0,x_0)} = R(x,y) \mu(x,x_0) \mu(y,x_0),
		\quad \mbox{with} \quad \mu(x,x_0) = \frac{x-x_0}{1-xx_0},
	\end{equation}
	which shows that $(f(\cdot)\rvert f(x_0) = 0)$ is a Gaussian process equal in distribution
	to $\mu(\cdot,x_0)f(\cdot)$. Using the linearity of the derivative, one gets \cite{MS13_supp}
	\begin{equation}\label{Pf2}
		(f'(x_1),f'(x_2),\ldots,f'(x_n)\rvert f(x_1)=f(x_2)=\ldots=f(x_m)=0) 
		\stackrel{d}{=} (M(x_1,\mathbf{x})f(x_1),M(x_2,\mathbf{x})f(x_2),\ldots,M(x_m,\mathbf{x})f(x_m)),
	\end{equation}
	where $M(\cdot,\mathbf{x}) = \left(\prod\limits_{i}\mu(\cdot,x_i)\right)'$ 	
	with the derivative taken with
	respect to the first argument
	and hence, one can check that $M(x_i,\mathbf{x})=\frac{1}{1-x_i^2}\prod\limits_{j\neq i} \frac{x_i-x_j}{(1-x_ix_j)^2}$.
	Combining Eqs. (\ref{Pf1}) and (\ref{Pf2}), one gets, after some manipulations \cite{MS13_supp}
	\begin{multline}\label{e:der_pfaff}
		\frac{\partial^{2n}}{\partial {x_1} \partial {x_2} \ldots \partial{x_{2n}}}
		S(x_1,x_2,\ldots,x_{2n}) = 
		\left(\frac{2}{\pi}\right)^{n}\frac{1}{\sqrt{\det R(x_i,x_j)}}		
		\prod\limits_{i} M(x_i,\mathbf{t}) 
		\E\left[f(x_1)f(x_2)\ldots f(x_{2n})\right]
		\\
		=\left(\frac{2}{\pi}\right)^{n}
		\prod\limits_{i< j} \sgn(x_j-x_i)
		\Pf\left(\mathbb{K}_{11}(x_i,x_j)\right)	
		= \left(\frac{2}{\pi}\right)^{n}
		\prod\limits_{i< j} \sgn(x_j-x_i)
		\frac{\partial^{2n}}{\partial {x_1} \partial {x_2} \ldots \partial{x_{2n}}}
		\Pf\left(\mathbb{K}_{22}(x_i,x_j)\right),
	\end{multline}
	where
	\begin{equation}\label{Pf4}
		\mathbb{K}\left(x,y\right) = 
		\begin{pmatrix}
			\mathbb{K}_{11}\left(x,y\right) & \mathbb{K}_{12}\left(x,y\right) \\
			\mathbb{K}_{12}\left(x,y\right) & \mathbb{K}_{22}\left(x,y\right)
		\end{pmatrix}
		=
		\begin{pmatrix}
			\frac{x-y}{\sqrt{1-x^2}\sqrt{1-y^2}(1-xy)^2} & \sqrt{\frac{1-y^2}{1-x^2}}\frac{1}{1-xy}\\
			-\sqrt{\frac{1-x^2}{1-y^2}}\frac{1}{1-xy} & \sgn(x-y)\arcsin \frac{\sqrt{1-x^2}\sqrt{1-y^2}}{1-xy}
		\end{pmatrix},
	\end{equation}	
	with $\mathbb{K}_{11}(x,y) = \frac{\partial^2}{\partial x\partial y}\mathbb{K}_{22}(x,y)$, $\mathbb{K}_{12}(x,y) = \frac{\partial}{\partial x}\mathbb{K}_{22}(x,y)$ and $\mathbb{K}_{21}(x,y) = \frac{\partial}{\partial y}\mathbb{K}_{22}(x,y)$. We refer the reader to Ref.~\cite{MS13_supp} for the details of the computations leading to the Pfaffian expression given in Eqs. (\ref{e:der_pfaff}) and (\ref{Pf4}). Finally, using appropriate boundary values of both sign-correlation function and corresponding Pfaffian 
	we can integrate over the $x_i$'s on both sides of \eqref{e:der_pfaff} to get that, for any ordered $2n$-tuples
	$-1<x_1<x_2<\ldots<x_{2m}<1$, one has
	\begin{equation}\label{pfaffian_correl_sup}
		S(x_1,x_2,\ldots,x_{2m}) = \left(\frac{2}{\pi}\right)^{m} \Pf(\mathbb{K}_{22}(x_i,x_j))\;.
	\end{equation}
We now use this identity (\ref{pfaffian_correl_sup}) to compute $\rho_m(x_1, \cdots, x_m)$ from Eqs. (\ref{e:MS_dens}) and (\ref{Pf1}). It is then convenient to relabel the variables and use Eq. (\ref{pfaffian_correl_sup}) with the relabelling $x_{2i + 1} \to x_i$ and $x_{2i} = s_i$. With this relabelling, we can now differentiate Eq. (\ref{pfaffian_correl_sup}) with respect to the variables of odd indices and then take the limit $x_{2i}\to
	x_{2i-1}$ for all $i=1,\ldots,m$, according to Eq. (\ref{Pf1}). This yields finally~\cite{MS13_supp}
	\begin{equation}\label{rho_n_Pfaffian}
		\rho_m(x_1, \cdots, x_m) = \pi^{-m}\Pf\left(\mathbb{K}(x_i,x_j)\right).
	\end{equation}
	
	We can now use this result (\ref{rho_n_Pfaffian}) to compute the multi-time sign-correlation function for the $2d$-diffusion field. Indeed, the $2d$-diffusion field $\varphi({\bf 0},t)$ coincides, for large time $t\gg1$, with the Kac polynomials $f(x)$ for $x$ close 1 \cite{SM07_supp, SM08_supp}. More precisely, if one considers the normalised process $X(t) =  {\varphi({\bf 0},t)}/{\langle \varphi^2({\bf 0},t)\rangle}$, one has
	\begin{eqnarray}\label{identity_diffusion_kac}
	X(t) = \frac{\varphi({\bf 0},t)}{\langle \varphi^2({\bf 0},t)\rangle} \overset{d}{\simeq} f(x=1-1/t) \;, \; t \to \infty \;.
	\end{eqnarray}
Therefore, from Eqs. (\ref{def_S}) and (\ref{pfaffian_correl_sup}), one obtains that the multi-time correlation functions 	of $\sgn X(t)$ also have a Pfaffian structure with kernel ${\mathbb K}_{22}(x_i = 1-1/t_i, x_j = 1-1/t_j)$ with $t_i, t_j \gg 1$. Finally, using that
\begin{eqnarray}\label{K_asympt}
{\mathbb K}_{22}(x_i = 1-1/t_i, x_j = 1-1/t_j) \approx \sgn(t_j - t_i) {\rm arcsin} \left( \frac{2 \sqrt{t_i\, t_j}}{t_i + t_j}\right) \;,\;\; t_i, t_j \gg 1 \;,
\end{eqnarray}
we obtain the formulae given Eqs. (20) and (21) in the text. In particular, specifying Eqs. (\ref{pfaffian_correl_sup}) and (\ref{K_asympt}) to $m=1$, we obtain
\begin{eqnarray}\label{2point_correl}
\langle X(t_1) X(t_2) \rangle \approx  \sgn(t_2 - t_1) \frac{2}{\pi}  {\rm arcsin} \left( \frac{2 \sqrt{t_1\, t_2}}{t_1 + t_2}\right) \;,\;\; t_1, t_2 \gg 1 \;.
\end{eqnarray}

	\section{6) Pfaffian structure of the GSP with correlator $c(T) = {\rm sech}(T/2)$}

	{
	In this section we extend the ideas of \cite{MS13_supp} to another process that plays a crucial role
	in our study (see Fig. 1 in the main text), namely the stationary GSP $Y(T)$ with correlation function $c(T) =  {\rm sech}(T/2)$ and  
	study the corresponding distribution of zeros. Similarly to \eqref{e:MS_dens}
	 we compute the correlation functions of the zeros by studying the conditional process 
	$(Y(\cdot)|Y(t)=0)$, which is also a GSP with correlation function (see Eq. (\ref{Rtilde}) in the previous section)
	\begin{equation*}
		c(x-y) - \frac{c(x-t)c(y-t)}{c(0)} = \sech\left(\frac{x-y}{2}\right) 
		- \sech\left(\frac{x-t}{2}\right)\sech\left(\frac{y-t}{2}\right) = c(x-y)\tanh(x-t)\tanh(y-t),
	\end{equation*}
where the last equality can be checked by using standard (hyperbolic) trigonometric relations. Therefore $(X(\cdot) \rvert X(t) =0) \stackrel{d}{=} \mu(\cdot,t) X(t)$ with $\mu(x,t) = 
	\tanh\frac{x-t}{2}$. Now using the linearity of the derivative we obtain (cf. \eqref{Pf2})
	\begin{equation*}
		(f'(t_1),f'(t_2),\ldots,f'(t_m)\rvert f(t_1)=f(t_2)=\ldots=f(t_m)=0) 
		\stackrel{d}{=} (M(t_1,\mathbf{t})f(t_1),M(t_2,\mathbf{t})f(t_2),\ldots,M(t_m,\mathbf{t})f(t_m)),
	\end{equation*}
	with $M(\cdot,\mathbf{t}) = \left(\prod\limits_{i}\mu(\cdot,t_i)\right)'$
	and $M(t_i,\mathbf{t})=\frac{1}{2}\prod\limits_{j\neq i} \tanh\frac{t_i-t_j}{2}$. {Using that $\frac{d}{dt}\sgn(t) = 2\delta (t)$, we obtain
	(for more details see \cite{MS13_supp} and references therein)
	\begin{eqnarray*}
		\frac{\partial^{2m}}{\partial {t_1} \partial {t_2} \ldots \partial_{t_{2m}}}
		\E\left[\sgn X\left(t_1\right)\ldots \sgn X\left(t_{2m}\right)\right]
		&=& 2^{2m} \E\left[\delta\left(X(t_1)\right) X'(t_1)\ldots \delta\left(X(t_{2m})\right) X'(t_{2m})\right]\\
		&=& \frac{2^m}{\pi^m \sqrt{\det c(t_i-t_j)}}
		\E\left[ X'(t_1)\ldots  X'(t_{2m})\vert X(t_1)=\ldots=X(t_{2m})=0\right],
	\end{eqnarray*}
	where, in the last equality, we used the explicit expression for the probability density
	of vector the $(X(t_i))$ at 0. Using the previous result on the conditional process $(X(.)\vert X(t)=0)$
	and the Wick theorem, we obtain (cf. \eqref{e:der_pfaff})
	\begin{eqnarray}
	\frac{\partial^{2m}}{\partial {t_1} \partial {t_2} \ldots \partial {t_{2m}}}
	\E\left[\sgn X\left(t_1\right)\sgn X\left(t_2\right)\ldots \sgn X\left(t_{2m}\right)\right] 
	&=&
	\left(\frac{2}{\pi}\right)^{m}\frac{1}{\sqrt{\det \sech\frac{t_i-t_j}{2}}}		
	\prod\limits_{i} M(t_i,\mathbf{t}) 
	\E\left[X(t_1)X(t_2)\ldots X(t_{2m})\right] \nonumber 
	\\
	&=&\left(-2\pi\right)^{-m}
	\frac{\prod\limits_{i<j}\tanh^2\frac{t_i-t_j}{2}}{\sqrt{\det\sech\frac{t_i-t_j}{2}}}
	\mathrm{Hf}\left(\sech\frac{t_i-t_j}{2}\right) \;, \label{formula_Hf}
	\end{eqnarray}
	where $\mathrm{Hf}(A)$, for a $2m \times 2m$ symmetric matrix $A = (a_{i,j})_{1 \leq i,j \leq m}$ is defined as $\mathrm{Hf}(A) =  1/(2^m m!)\sum_{\sigma \in {{\cal S}_{2m}}} \prod_{i=1}^m a_{\sigma(2i-1), \sigma(2i)}$, where ${\cal S}_{2m}$ is the group of permutations of $2m$ elements. The denominator in the above formula can be evaluated by using the following formula for a Cauchy's determinant
	\begin{equation*}
		\det \left(\frac{1}{1-x_iy_j}\right)_{i,j=1}^k = 
		\frac{\prod\limits_{1\leq i<j\leq k}\left(x_i-x_j\right)\left(y_i-y_j\right)}{
		\prod\limits_{1\leq i,j\leq k}\left(1-x_iy_j\right)},
	\end{equation*}
	with $x_i=e^{t_i}$ and $y_j=-e^{-t_j}$. We then obtain
	\begin{equation*}
		\det \sech \frac{t_i-t_j}{2} = 2^{2m} \det\frac{1}{1+e^{t_i-t_j}} = 2^{2m}
		\frac{\prod\limits_{1\leq i< j\leq 2m}4\sinh^2 \frac{t_i-t_j}{2}}{
			2^{2m} \prod\limits_{1\leq	i<j\leq 2m}{4\cosh^2\frac{t_i-t_j}{2}}}
		= \prod\limits_{1\leq i<j\leq 2m} \tanh^2\frac{t_i-t_j}{2},
	\end{equation*}
	and after combining with Eq. (\ref{formula_Hf}), we get
	\begin{eqnarray*}
	\frac{\partial^{2m}}{\partial {t_1} \partial {t_2} \ldots \partial {t_{2m}}}
	\E\left[\sgn X\left(t_1\right)\sgn X\left(t_2\right)\ldots \sgn X\left(t_{2m}\right)\right] 
	&=&
	\left(2\pi\right)^{-m}\prod\limits_{1\leq i<j\leq 2m}	
	\sgn(t_j-t_i) \tanh\frac{t_i-t_j}{2}
	\mathrm{Hf}\left(\sech\frac{t_i-t_j}{2}\right).
	\end{eqnarray*}
	Using a formula due to Ishikawa, Kawamuko, and Okada \cite{IKO05_supp}
	\begin{equation}\label{IKO}
		\prod\limits_{1\leq i < j\leq 2n} \frac{x_i-x_j}{x_i+x_j}
		\mathrm{Hf}\left(\frac{1}{x_i+x_j}\right)_{i,j=1}^{2n} =
		\Pf\left(\frac{x_i-x_j}{(x_i+x_j)^2}\right)_{i,j=1}^{2n},		
	\end{equation}
	we obtain the derivative of the multi-time spin-correlation function under as a Pfaffian. Indeed, by taking $x_i= e^{t_i}$
	we obtain
	\begin{equation*}
		\frac{\partial^{2m}}{\partial {t_1} \partial {t_2} \ldots \partial {t_{2m}}}
		\E\left[\sgn X\left(t_1\right)\ldots \sgn X\left(t_{2m}\right)\right]
		= (2\pi)^{-m}\prod\limits_{i<j}
		\sgn(t_j-t_i) \Pf\left(\frac{\sinh\frac{t_i-t_j}{2}}{\cosh^2\frac{t_i-t_j}{2}}\right).
	\end{equation*}
	Spin correlation function can now be obtained by direct integration. As a boundary condition 
	one should use decay of	correlation functions at infinity. All together this implies that
	for $t_1<t_2<\ldots<t_{2m}$
	\begin{equation*}
		\E\left[\sgn X\left(t_1\right)\sgn X\left(t_2\right)\ldots \sgn X\left(t_{2m}\right)\right] 
		 = \left(2\pi\right)^{-m}
		\Pf\left(8\arctan\tanh\frac{t_i-t_j}{4}-2\pi\sgn\left(t_i-t_j\right)\right).
	\end{equation*}
	}
	Performing differentiation and taking limits we finally obtain
	\begin{equation}
		\rho_m\left(t_1,t_2,\ldots,t_m\right) = \pi^{-m} \Pf
		\begin{pmatrix}
			\frac{1}{4}
			\frac{\sinh\frac{t_i-t_j}{2}}{\cosh^2\frac{t_i-t_j}{2}} & 
			\frac{1}{2}\sech\frac{t_i-t_j}{2}\\
			- \frac{1}{2}\sech\frac{t_i-t_j}{2} & 2\arctan\tanh\frac{t_i-t_j}{4}-\frac{1}{2}\pi\sgn(t_i-t_j)
		\end{pmatrix} \;,
	\end{equation}
}
with the convention that $\sgn(0) = 0$.

\section{7) Glauber dynamics for the Ising model on a half line and mapping to the $2d$-diffusion equation}

We consider a semi-infinite Ising spin chain, whose configuration at time $t$ is given by 
$\{\sigma_i(t)\}_{i \geq 0}$, with $\sigma_i(t) = \pm 1$. Initially, the system is in a 
random initial configuration where $\sigma_i(0) = \pm 1$ with equal probability $1/2$ and,
at subsequent time, the system evolves according to the Glauber dynamics. Within each 
infinitesimal time interval $\Delta t$, every spin is updated according to 
\begin{eqnarray}\label{e:Glauber_supp}
\sigma_i(t+ \Delta t) = 
	\begin{cases}
		&\sigma_i(t) \;, \; \hspace*{0.5cm}{\rm with \; proba. \;} 1 - 2 \Delta t, \; \\
		&\sigma_{i-1}(t)  \;, \; \hspace*{0.15cm}{\rm with \; proba. \;} \Delta t, \\
		&\sigma_{i+1}(t)  \;, \; \hspace*{0.2cm}{\rm with \; proba. \;} \Delta t,
	\end{cases}
\end{eqnarray} 
while the site at the $\sigma_0(t)$ evolves via $\sigma_{0}(t + \Delta t) = \sigma_0(t)$ with 
probability $1 - \Delta t$ and $\sigma_0(t) = \sigma_1(t)$ with probability 
$\Delta t$.
\begin{figure}
\includegraphics[width=0.7\linewidth]{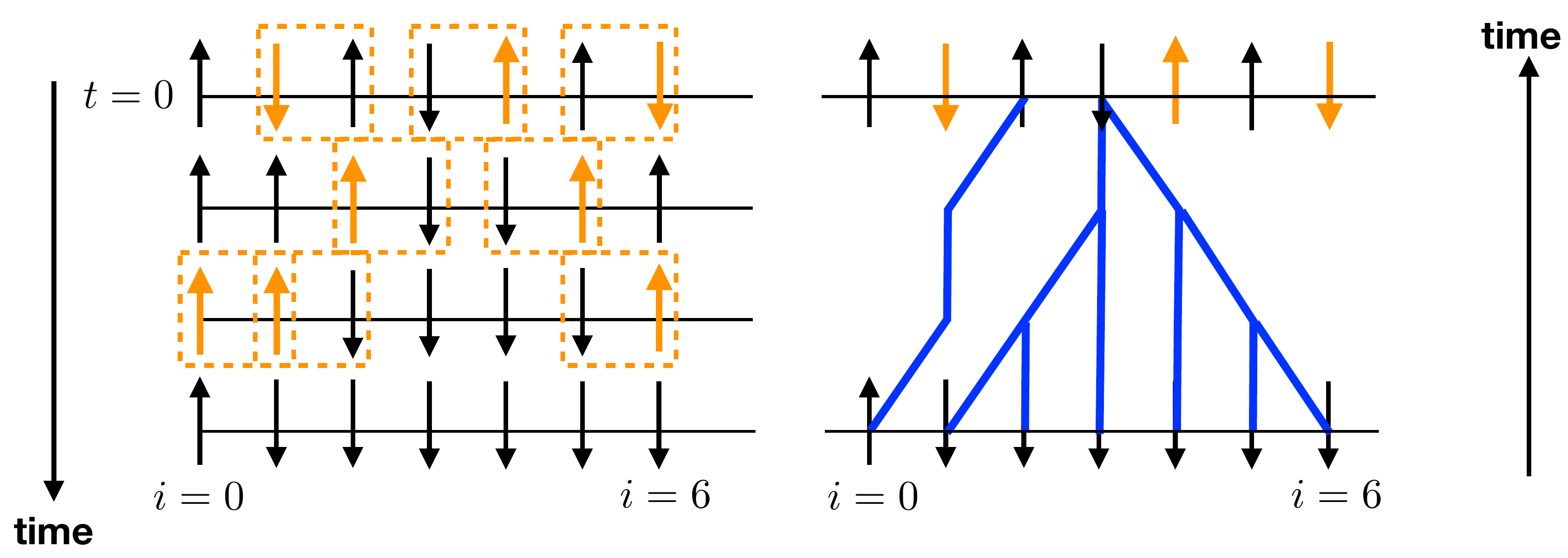}
\caption{Illustration of the mapping between the Glauber dynamics (\ref{e:Glauber_supp}), in the left panel, and coalescing paths, in the right panel. {\bf Left panel}: at each time steps, the state of the orange spins are changed (while the black spins stay the same) and they take the value of the neighbouring spin inside the square box. {\bf Right panel:} the arrow of time is reversed such that the initial configuration of the walks is actually the final state of the spin system. At each time step a given path connects the sites $i$ and $j$ (with $j=i-1, i$ or $i+1$) such that $\sigma_j$ is the ``ancestor'' of $\sigma_i$. For instance, during the final step in the left panel, the spin at site $i=0$ took the value of the neighbouring spin at site $i=1$, hence the first step of the walk that starts at $i=0$ goes from $i=0$ to $i=1$.}\label{Fig:mapping}
\end{figure}

Derrida et al. in \cite{DHP95_supp,DHP96_supp} noticed that if one traces back in time
the value of the spin $i$ at time $t$, one obtains a random walk that connects the site
$i$ at time $t$ through its various ancestors to a particular site $i_0$ at initial time $t=0$ (see Fig. \ref{Fig:mapping}). And consequently
$\sigma_i(t) = \sigma_{i_0}(0)$. It is important to notice that, in this mapping, the arrow of time has to be reversed (see Fig. \ref{Fig:mapping}). Then, for instance, to compare the values of the spin $\sigma_i(t)$ and $\sigma_j(t)$
at two different sites $i$ and $j$, one considers two random walkers starting at site $i$ and $j$. After time $t$, two 
different situations may then have occurred: (i) either the two random walkers have ``coalesced'' into a single
walker that ``ends up'' on site $k_0$, and consequently $\sigma_i(t) = \sigma_j(t) = \sigma_{k_0}(0)$, (ii) or the random
walkers have not met each other and therefore $\sigma_{i}(t) = \sigma_{i_0}(t)$ while $\sigma_j(t) = \sigma_{j_0}(t)$ with
$i_0 \neq j_0$. Using this mapping to coalescing random walkers, the authors of  \cite{DHP95_supp,DHP96_supp} computed the
persistence probability $p_{\rm Ising}(t)$ for the Ising chain and found that, for large $t \gg 1$, $p_{\rm Ising}(t) \sim t^{-\theta_{\rm Ising}}$ with
$\theta_{\rm Ising}= 3/8$. They further generalized this result to the $q$-states Potts model ($q=2$ corresponding to the Ising model) and found
that the persistence probability $p_{\rm Potts}(t)$ decays as~\cite{DHP95_supp,DHP96_supp}   
\begin{equation}\label{Potts}
p_{\rm Potts}(t)	 \sim t^{-\widehat{\theta}(q)}  \quad , \quad 
	\widehat{\theta}(q) = -\frac{1}{8} + \frac{2}{\pi^2}\left[\cos^{-1}\frac{2-q}{\sqrt{2}q}\right]^2.
\end{equation}
Remarkably, the expression for $\widehat\theta(q)$ in (\ref{Potts}) bears strong similarities with the function $\psi(s)$, associate to the 
$2d$-diffusing field given in the text in Eq. (19). Below we show that this is not just a coincidence: indeed we exhibit a mapping between the
(semi-infinite) Ising-chain with Glauber dynamics and the $2d$-diffusion equation with random initial conditions. Using the formulation of the Glauber 
dynamics \eqref{e:Glauber_supp} in terms of coalescing random walks, that the multi-time correlation functions of 
$\sigma_0$ are given by a Pfaffian, as given in Eq. \eqref{Pfaffian_sigma} in the text, 
for $t_1< t_2< \cdots< t_{2m}$
\begin{eqnarray}\label{e:Pfaffian_sigma0}
\langle \sigma_0(t_1) \cdots \sigma_0(t_{2m})\rangle \sim {\rm Pf}(A) \;, 
\quad a_{i,j} = \sgn(j-i)\langle \sigma_0(t_i)\sigma_0(t_j)\rangle
\end{eqnarray} 
with $a_{i,j}$ as given in Eq. (21) in the text for 
$t_i,t_j \gg 1$. From these 
identities for any correlation function in Eqs. \eqref{Pfaffian_sigma} and \eqref{e:Pfaffian_sigma0}, 
we conclude that $\sgn(X(t))$ for the $2d$-diffusion equation and $\sigma_0(t)$ in the 
semi-inifinite Ising chain with Glauber dynamics are actually exactly the same process 
in the large time limit. Therefore we conclude that their persistence properties do coincide 
and therefore $b = 3/16$, as announced in Eq. \eqref{exact_theta_2}.

\begin{figure}
\includegraphics[width = 0.4\linewidth]{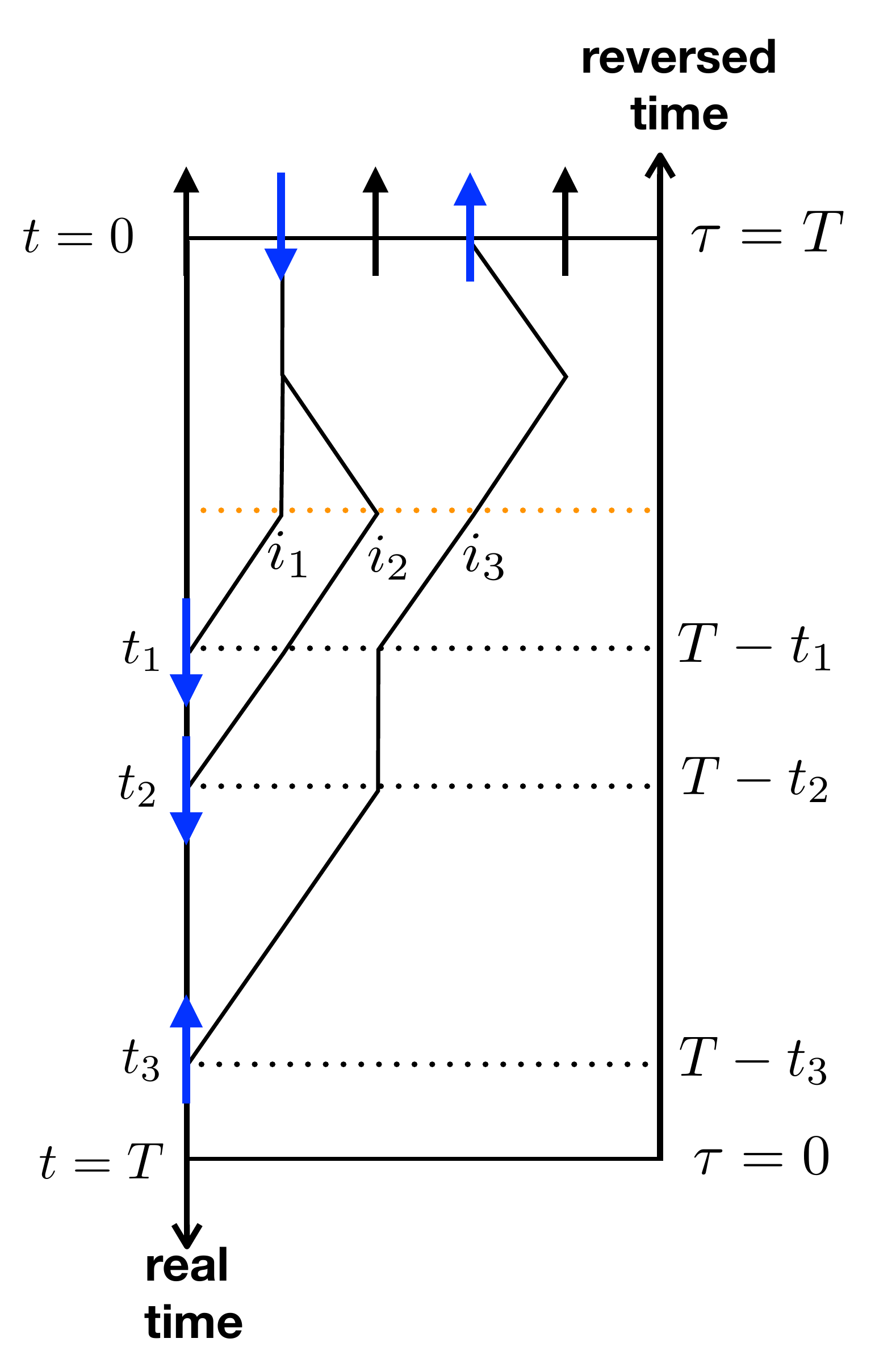}
\caption{Illustration of the mapping to coalescing random walks to compute the multi-time correlation functions of the spin at the origin at $N=3$ different times $t_1, t_2$ and $t_3$ . In this specific realisation of the Glauber dynamics with a given initial configuration at $t=0$, one reads immediately $\sigma_0(t_1) \sigma_0(t_2) \sigma_0(t_3) = (\sigma_1(0))^2\, \sigma_4(0)$. Note that, at all time, $i_1 \leq i_2 \leq i_3$.}\label{Fig:coalesc_t0}
\end{figure}

To compute these multiple-time correlation functions $\langle \sigma_0(t_1) \cdots \sigma_0(t_{N})\rangle$ we again use the mapping of the $T=0$ Glauber dynamics to coalescing random walks (see Fig. \ref{Fig:coalesc_t0}). A remarkable simplification occurs when one considers the site at the boundary $i=0$ of the semi-infinite~\cite{DHP95_supp, DHP96_supp}. Indeed, when $N$ walkers start from the origin at times $T - t_N< T -t_{N-1} < \cdots < T - t_1$ the positions of the path at any time $t$ remain always in the same order, i.e. $i_1\leq i_2 \leq i_N$, since when a walker is at the origin, it can only hop to the right (see Fig. \ref{Fig:coalesc_t0}). 
As we will see, the main tool for computing the multi-time correlation functions is the probability $c_{i,j}$ that two walkers, starting at $T-t_i > T-t_j$, with $i < j$, do not meet up to time $T$. This probability $c_{i,j}$ is of course independent of $T$ and reads, for $t_j> t_i \gg 1$ with $t_i/t_j$ fixed, \cite{DHP95_supp, DHP96_supp}
\begin{equation}\label{cis}
	c_{i,j} \simeq \frac{4}{\pi}\tan^{-1}\sqrt{\frac{t_j}{t_i}} - 1 \;, \; {\rm for} \; \; t_j > t_i \gg 1 \;.
\end{equation}
Similarly, if $1 \ll t_j<t_i$, one finds~\cite{DHP95_supp, DHP96_supp}
\begin{eqnarray}\label{cis_inv}
c_{i,j} = - c_{j,i} \;, \;  {\rm for} \; \; t_i > t_j \gg 1 \;.
\end{eqnarray}
In Ref. \cite{DHP95_supp, DHP96_supp}, this result (\ref{cis}) was generalized to compute the propability $c^{\left(k\right)}_{j_1,j_2,\ldots,j_{2k}}$ that no pair of random walks (among the $2k$ random walkers) with labels $j_1,j_2,\ldots,j_{2k}$ meets up to time $T$. It turns out that it can be written as a Pfaffian, 
\begin{equation}\label{e:non_intersect}
c^{\left(k\right)}_{j_1,j_2,\ldots,j_{2k}} = \Pf\left\{c_{i,j}\right\}_{i,j = j_1,\ldots,j_{2k}}  = \frac{1}{k!\,2^k} \sum_{\sigma \in {\cal S}_{2k}} \epsilon(\sigma) c_{j_{\sigma(1)}, j_{\sigma(2)}} \cdots c_{j_{\sigma(2k-1)}, j_{\sigma(2k)}} \;,
\end{equation}
where the sum is over the group of permutations ${\cal S}_{2k}$ of the indices $\{j_1, j_2, \cdots, j_{2k}\}$ and $c_{i,j}$ is the anti-symmetric matrix defined in Eqs. (\ref{cis}) and (\ref{cis_inv}).

Let us start with the two-point correlation function $\langle \sigma_0(t_i)\sigma_0(t_j) \rangle$ which can easily be computed using the mapping to coalescing random walks (see Fig. \ref{Fig:coalesc_t0}). If, after time $t$ the two paths, starting at $t-t_i$ and $t-t_j$ have not coalesced, which happens with probability $c_{i,j}$ then $\sigma_0(t_i) \sigma_0(t_j) = \sigma_{j_0}(0) \sigma_{k_0}(0)$ for some  $j_0 \neq k_0$. Thus, after averaging over the initial conditions $\sigma_{j_0}(0) = \pm 1$, as well as $\sigma_{k_0}(0) = \pm 1$, with equal probability, we see that such non-coalescing paths give a vanishing contribution to the two-point correlation function $\langle \sigma_0(t_i) \sigma_0(t_j) \rangle$. On the contrary, if the two paths have coalesced, which happens with probability $1-c_{i,j}$, then there exists a site $j_0$ such that $\sigma_0(t_i) \sigma_0(t_j) = (\sigma_{j_0}(0))^2 = 1$, which remains $1$ after averaging over the initial condition $\sigma_{j_0}(0) = \pm 1$. Hence, such coalescing paths give a contribution $1-c_{i,j}$ to the two-time correlation function and therefore, for $t_j> t_i \gg 1$ with $t_i/t_j$ fixed, one has 
\begin{equation}\label{correl_atan}
	\langle \sigma_0(t_i)\sigma_0(t_j)\rangle = 1-c_{i,j} \simeq 2 - \frac{4}{\pi} \tan^{-1}\sqrt{\frac{t_j}{t_i}} = \frac{4}{\pi} \tan^{-1}\sqrt{\frac{t_i}{t_j}}   \;,
\end{equation}
where, in the last equality, we have used $\tan^{-1}(x) = \pi/2 - \tan^{-1}(1/x)$. Using the identity
\begin{eqnarray}\label{identity}
\tan^{-1}(x) = \frac{1}{2} \, \sin^{-1} \left(\frac{2\,x}{1+x^2} \right) \;, \;\; {\rm for}\;\;\; 0<  x < 1 \;,
\end{eqnarray}
we obtain from (\ref{correl_atan}) that the two-time correlation function $\langle \sigma_0(t_i)\sigma_0(t_j)\rangle$ reads, for $t_j>t_i \gg 1$ 
\begin{eqnarray}\label{correl_a}
\langle \sigma_0(t_i)\sigma_0(t_j)\rangle \simeq \frac{2}{\pi} \sin^{-1} \left( \frac{2 \sqrt{t_i\, t_j} } {t_i + t_j}\right) \;.
\end{eqnarray}
Similarly, using the relation in (\ref{cis_inv}), we obtain, for $t_i, t_j \gg 1$,
\begin{eqnarray}\label{correl_complete}
\langle \sigma_0(t_i)\sigma_0(t_j)\rangle \simeq {\rm sign}(j-i)\frac{2}{\pi} \sin^{-1} \left( \frac{2 \sqrt{t_i\, t_j} } {t_i + t_j}\right) \;,
\end{eqnarray}
which thus coincides precisely with the $2$-point correlation of $\sgn(X(t)))$ given in Eq. (\ref{2point_correl}), with $X(t)$ the normalised $2d$-diffusing field.

\begin{figure}
	\includegraphics[width = 0.6\linewidth]{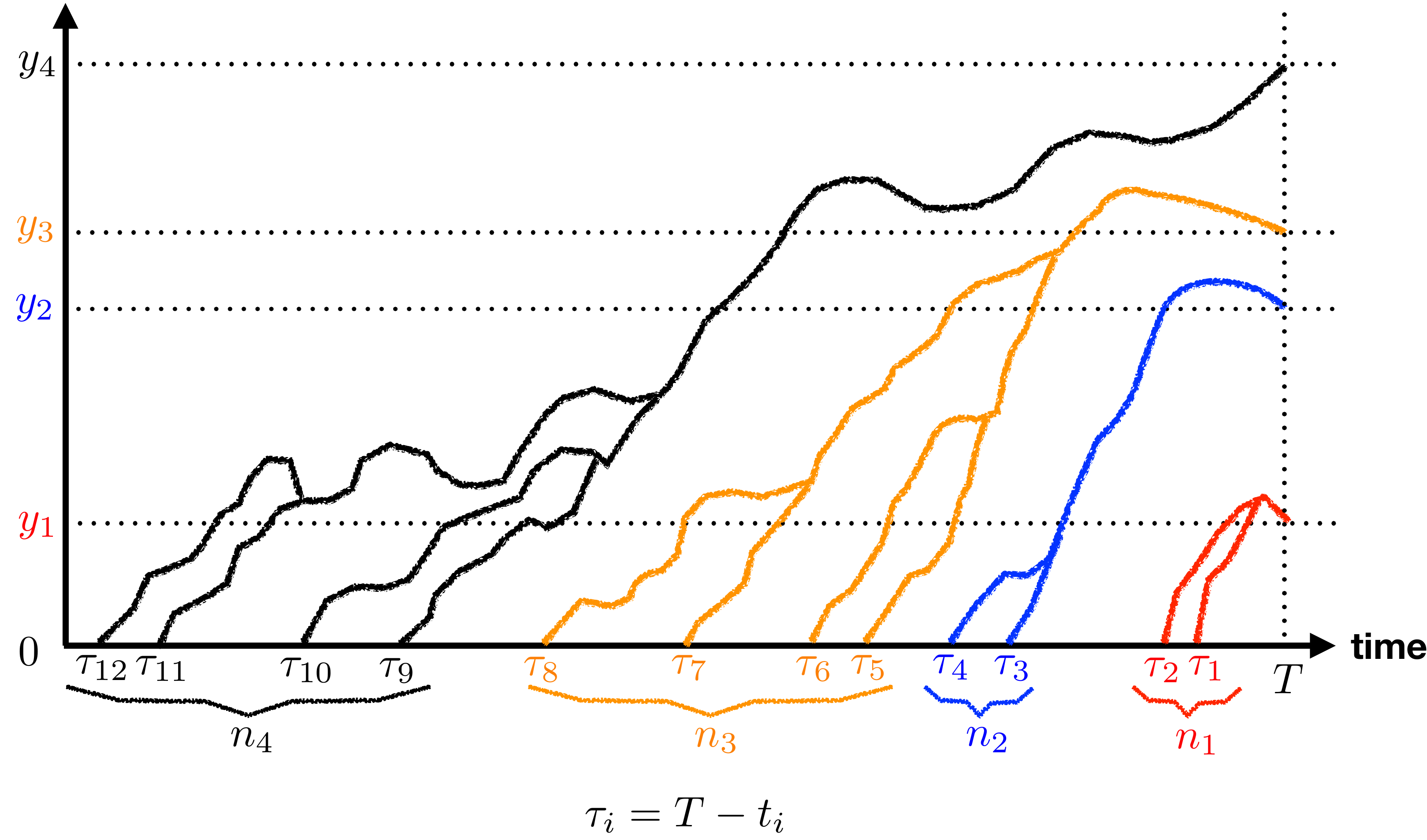}
	\caption{Sample paths configuration for $2m=12$ random walkers ending up in $m' = 4$ distinct points
	$y_1<y_2<y_3<y_4$ and coalesced into $4$ groups of sizes $n_1=2, n_2=2, n_3=4, n_4=4$. 
	This configuration belongs to the set $\mathcal{A}_{\mathbf{m}}$ for partition $\mathbf{m} = (2,2,4,4)$.
	The partition $\mathbf{m}$ complies with non-intersection events $c^{(k)}_{j_1,j_2,\ldots,j_{2k}}$ 
	such that all $(j_1,j_2,\ldots,j_{2k})$ belong to different intervals chosen from $\Delta_1=\left[1,2\right], 
	\Delta_2=\left[3,4\right], \Delta_3=\left[5,8\right], \Delta_4=\left[9,12\right]$. For example, $c^{(2)}_{1,3}, c^{(2)}_{4,11}$, 
	$c^{(4)}_{1,3,5,9}$, etc.}
	\label{Fig:supp_coalesc}
\end{figure}

We now prove relation \eqref{e:Pfaffian_sigma0} for any $m \geq 1$. Let us denote $T<t_{2m} < \ldots < t_{1}$ and use
the mapping to coalescing random walks in $\Z_+$ starting from 
the origin at times $T-t_1>T-t_2>\ldots>T-t_{2m}$ (see Fig. \ref{Fig:coalesc_t0}). As explained in the case $m=1$, 
if two of the random walks meet before time $T$, the corresponding spins have the same ancestor and therefore the same value. On the contrary, if they do not meet, the values of the 
spins take independently different values $\pm 1$ with probabilities $1/2$. Assume then that the RWs end up at the points
$x_1\leq x_2\leq \ldots \leq x_{2m}$. For every partition $\mathbf{m} = \left(n_1,n_2,\ldots,n_{m'}\right)$ of 
$2m = n_1 + n_2 + \cdots + n_{m'}$ we introduce the event ${\cal A}_{\mathbf{m}}$ consisting of paths satisfying
\begin{equation}\label{def_Am}
{\cal A}_{\mathbf{m}} = \left\{\mbox{configurations of coalescing RW with } 
x_1=\ldots=x_{n_1} < x_{n_1+1}=\ldots=x_{n_1+n_2} < \ldots < x_{2n-n_{m'}+1}=\ldots=x_{2m}	
\right\}.
\end{equation}
We denote by $y_1,y_2,\ldots,y_{m'}$ the distinct points among $x_1,x_2,\ldots,x_{2m}$ (see Fig. 
\ref{Fig:supp_coalesc}).
It is useful to introduce
$\Delta_1,\ldots,\Delta_{m'}$ which are the corresponding intervals of indices, i.e.,
\begin{eqnarray} \label{def_delta}
\Delta_1 = [1,n_1] \;, \; \Delta_{\ell} = \left[n_1+\ldots+n_{\ell-1}+1,n_1+\ldots+n_{\ell}\right] \;{\rm for \;} \ell \geq 1 \;,
\end{eqnarray}
as well as $N_1=1,N_2 = n_1+1,\ldots,N_{m'} = n_1+\cdots+ n_{m'-1}+1$ which are the left endpoints of these intervals $\Delta_1,\Delta_2,\ldots,\Delta_{m'}$.
The only configurations that will contribute to the expectation in the l.h.s. of \eqref{e:Pfaffian_sigma0} are such that all $n_i$ are even. Indeed,
\begin{equation}\label{formula1}
\E\left[\sigma_0(t_1) \sigma_0(t_2) \cdots \sigma_0(t_{2m})\rvert {\cal A}_{\mathbf{m}}\right]= 
\E\left[\sigma_{x_1}(0) \sigma_{x_2}(0) \cdots \sigma_{x_{2m}}(0)\rvert {\cal A}_\mathbf{m}\right]
= \Pr\left[{\cal A}_\mathbf{m}\right]\prod\limits_{j=1}^{m'}\E_0\left[\sigma_{y_j}^{n_j}\left(0\right)\right]
= \Pr\left[{\cal A}_\mathbf{m}\right] {\mathbb I}_{2 \mid n_1,\ldots,n_{m'}},
\end{equation}
where ${\mathbb I}_{2 \mid n_1,\ldots,n_{m'}}$ imposes that all the $n_i$'s are even. In Eq. (\ref{formula1}), we have 
used the independence between the spins at time zero and denoted by $\E_0$ the average over the initial condition. 
Therefore, one can now rewrite multi-time spin autocorrelation function as follows
\begin{equation}\label{e:spin_autocorr}
\langle \sigma_0(t_1) \sigma_0(t_2) \cdots \sigma_0(t_{2m}) \rangle
= \sum_{\bf m} \E\left[\sigma_0(t_1) \sigma_0(t_2) \cdots \sigma_0(t_{2m})\rvert {\cal A}_{\mathbf{m}}\right] = \sum\limits_{\mathbf{m}\rvert n_i\in 2\N}
\Pr\left[{\cal A}_\mathbf{m}\right].
\end{equation}

The next step is to expand the Pfaffian in the right hand side of \eqref{e:Pfaffian_sigma0} by using the decomposition formula valid for any anti-symmetric matrices $A,B$ of the same size $2m \times 2m$ (see e.g. \cite{TZ2011_supp})
\begin{equation}\label{pfaffian_identity}
\Pf\left(A+B\right) = \sum\limits_{{\mathbf{J}}\subset\left\{1,2,\ldots,2m\right\}}
\left(-1\right)^{\sum j_k - \left|\mathbf{J}\right|/2}\Pf\left(A\vert_{\mathbf{J}}\right)\Pf\left(B\vert_{\mathbf{J}^c}\right),
\end{equation}
{where the sum runs over all even sized subsets $\mathbf{J} = (j_1, j_2, \cdots, j_{|\mathbf{J}|})$ with $|\mathbf{J}|$ elements ($\mathbf{J}^c$ denoting its complementary in the set $\{1,2,\cdots,2m \}$) and $A|_{\mathbf{J}}$ denotes a minor of the matrix $A$ containing rows and columns
with indices taken from $\mathbf{J}$ (and the Pfaffian of the empty matrix is taken to have value 1). By specializing this formula (\ref{pfaffian_identity}) to the matrices $A = \left\{\sgn\left(j-i\right)\right\}_{i,j=1}^{2m}$ (with the convention that $\sgn(0) = 0$) and $B=- \left\{c_{i,j}\right\}_{i,j=1}^{2m}$, with $c_{i,j}$ as defined in Eqs. (\ref{cis}) and (\ref{cis_inv}) we get
\begin{eqnarray}
	\Pf\left(\sgn\left(j-i\right)-c_{i,j}\right) &=& 
	\sum\limits_{k=0}^m\sum\limits_{1\leq j_1<j_2<\ldots<j_{2k}\leq 2m}
	\left(-1\right)^{j_1+j_2+\ldots+j_{2k}}\Pf\left\{c_{i,j}\right\}_{j_1,j_2,\ldots,j_{2k}} \\
&=& \sum\limits_{k=0}^m\sum\limits_{1\leq j_1<j_2<\ldots<j_{2k}\leq 2m}
	\left(-1\right)^{j_1+j_2+\ldots+j_{2k}} c^{\left(k\right)}_{j_1,j_2,\ldots,j_{2k}}	 \label{pf_supp1}
\end{eqnarray}
where, in the second line, we have used that the Pfaffian appearing in the first line has a probabilistic interpretation in terms of non-intersecting paths [see Eq. (\ref{e:non_intersect})]. Note that when applying the formula (\ref{pfaffian_identity}) to the aforementioned matrices $A$ and $B$ we have used that $\Pf(A|\mathbf{J}) = 1$ for all even sized subset $\mathbf{J}$ as well as $\Pf(B|\mathbf{J}^c) = (-1)^{|\mathbf{J}^c|} \Pf(-B|\mathbf{J}^c)$. 

We recall that the event ${\cal A}_{\mathbf{m}}$ with partition $\mathbf{m} = \left(n_1,n_2,\ldots,n_{m'}\right)$ of $2m = n_1 + \cdots + n_{m'}$
consists of paths that meet at $m'$ distinct points $y_1,y_2,\ldots,y_{m'}$ at time $T$ [see Eq. (\ref{def_Am}), Fig. \ref{Fig:supp_coalesc}]. Therefore the probability of this event is the sum of all probabilities 
$c^{\left(k\right)}_{j_1,j_2,\ldots,j_{2k}}$ with the constraint that the indices $j_1< j_2< \ldots < j_{2k}$
belong to different intervals $\Delta_\ell$ of the partition (\ref{def_delta}). In this case we will say that
$\mathbf{J}$ comply with $\mathbf{m}$ and write $\mathbf{J}\sim \mathbf{m}$. We can thus rewrite the right hand side of Eq. (\ref{pf_supp1}) as
\begin{eqnarray}\label{pf_supp2}
	\Pf\left(\sgn\left(j-i\right)-c_{i,j}\right) = \sum_{\mathbf{m}} \alpha_{\mathbf{m}} \, {\rm Pr}[{\cal A}_m] \:, \; \alpha_{\mathbf{m}}=
\sum\limits_{k=0}^m\sum\limits_{\mathbf{J}\sim\mathbf{m}}\left(-1\right)^{j_1+j_2+\ldots+j_{2k}} \;.
\end{eqnarray}	
If $\mathbf{J}\sim\mathbf{m}$ then the index $j_1$ belongs to an interval $\Delta_{p_1}$, for some $p_1$. Therefore, all the subsets $\mathbf{J}' = \left\{j_1',j_2,\ldots, j_{2k}\right\}$ which differ from $\mathbf{J}$ only by the first index $j'_1$ such that $j'_1 \in \Delta_{p_1}$ also complies with the partition $\mathbf{m}$. By summing over such groups of subsets, i.e. over $j_1$, gives
\begin{equation}
\Sigma_{p_1}:= \sum\limits_{j_1 \in \Delta_{p_1}}\left(-1\right)^{j_1} = 
\begin{cases}
0, & {\rm if} \; n_{p_1} \; {\textrm{is even}}, \\
\left(-1\right)^{N_{p_1}}, & {\rm if} \; n_{p_1} \; {\textrm{is odd}}\;.
\end{cases} 
\end{equation}
Grouping now the subsets $\mathbf{J}$ that differ by two indices, and then by three indices, etc, we finally get
\begin{equation}\label{e:alpha_m}
\alpha_{\mathbf{m}} = \sum\limits_{k=0}^{[m'/2]}\sum\limits_{1\leq p_1<p_2<
	\ldots<p_{2k}\leq m'}\;\;\prod\limits_{\ell=1}^{2k} \Sigma_{p_\ell},
\end{equation}
where the empty product corresponding to $k=0$ is assumed to be equal to $1$ and where $[x]$ means the integer part of $x$. {The identity 
\eqref{e:alpha_m} means  
that the coefficient $\alpha_{\mathbf{m}}$ is equal to the sum of all even products of 
variables $\Sigma_1, \Sigma_2,\ldots,\Sigma_{m'}$. This can be easily rewritten as}
\begin{equation}
\alpha_{\mathbf{m}} = \dfrac{1}{2}\left[\prod\limits_{\ell=1}^{m'}\left(1+\Sigma_{\ell}\right)
+\prod\limits_{\ell=1}^{m'}\left(1-\Sigma_{\ell}\right)
\right].
\end{equation}
If for some $\mathbf{m}$ all $n_i$'s are even then for any
$\ell$ we have $\Sigma_{\ell} = 0$ and consequently $\alpha_{\mathbf{m}} = 1$.
For partitions containing odd intervals we denote $\ell_1$ and $\ell_2$ to be numbers
of the first two intervals of odd length. Then $N_{\ell_1}$ is odd and $\Sigma_{\ell_1} = -1$
and the first term in above expression vanishes. But one can also see that 
$N_{\ell_2}$ is even and $\Sigma_{\ell_2} = 1$ which implies that the second term vanishes as well. 
Therefore we see that $\alpha_{\mathbf{m}}\neq 0$ if and only if all $n_i$ are even. And hence 
the Pfaffian in the left hand side of Eq. (\ref{pf_supp2}) can be written as
\begin{eqnarray}\label{Pf_final}
\Pf\left(\sgn\left(j-i\right)-c_{i,j}\right) = \sum\limits_{\mathbf{m}\rvert n_i\in 2\N}  
\Pr\left[{\cal A}_\mathbf{m}\right]  = \langle \sigma_0(t_1) \sigma_0(t_2) \cdots \sigma_0(t_{2m}) \rangle\;,
\end{eqnarray}
where, in the last equality, we have used Eq. (\ref{e:spin_autocorr}). Finally, using the asymptotic behaviour of $1 - c_{i,j}$ for $t_i< t_j$ in Eqs. (\ref{correl_atan}) and (\ref{correl_a}), we obtain the result announced in Eq. (22) in the text.


\begin{thebibliography}{100}
\bibitem{footnote_pfaffian}
		We recall that, for an anti-symmetric matrix $A$ of size $2m \times 2m$, the Pfaffian is defined as ${\rm Pf}(A) =  1/(2^m m!)\sum_{\sigma \in {{\cal S}_{2m}}} \epsilon(\sigma) \prod_{i=1}^m a_{\sigma(2i-1), \sigma(2i)}$, where ${\cal S}_{2m}$ is the group of permutations of $2m$ elements.
\bibitem{footnote_continuation} Note that the computation of $\varphi(x)$ for $x>1/\pi$ requires an analytical continuation of $\psi(s)$ for $s>0$, which can be done straightforwardly from its explicit expression in Eq. (\ref{e:psi}).
\bibitem{footnote_clipped}
              The ``clipped'' process has obviously the same zero-crossings properties as $X(t)$ itself.
\bibitem{foot_smooth}
		Note that $\sigma_0$ is thus a smooth process, which is at variance with $\sigma_i$ with $i \gg 1$ which is non smooth.
\end{thebibliography}
\end{document}